\documentclass[5p,twocolumn]{elsarticle}

\usepackage{lineno,hyperref}
\usepackage{graphicx}
\usepackage{amsmath,amssymb,amsfonts}
\usepackage{bm}
\usepackage{algorithm}
\usepackage{algpseudocode}
\usepackage{array}
\usepackage{multirow}
\usepackage{enumitem}
\usepackage[caption=false,font=footnotesize]{subfig}
\usepackage{url}

\usepackage[final]{changes}

\newcolumntype{C}{>{\centering\arraybackslash}p{0.2\textwidth}}

\newcommand{\argmax}[1]{\underset{#1}{\operatorname{arg}\,\operatorname{max}}\;}

\modulolinenumbers[5]

\journal{Journal of Network and Computer Applications}

\begin{document}

\begin{frontmatter}

\fbox{\begin{minipage}[b][1cm][c]{18cm}
\footnotesize This article has been accepted for publication in the Journal of Network and Computer Applications, Elsevier. This is the author's version which has not been fully edited and content may change prior to final publication. Citation information: \url{https://doi.org/10.1016/j.jnca.2021.103020}
\end{minipage}}

\title{Application-Aware Resource Allocation and Data Management for MEC-assisted IoT Service Providers}

\author[1]{Simone Bolettieri}
\ead{s.bolettieri@iit.cnr.it}
\author[1]{Raffaele Bruno\corref{cor1}}
\ead{r.bruno@iit.cnr.it}
\author[2]{Enzo Mingozzi}
\ead{enzo.mingozzi@unipi.it}

\address[1]{IIT-CNR, Via G. Moruzzi 1, 56124 Pisa, ITALY}
\address[2]{Dept. of Information Engineering, University of Pisa, L.go L.Lazzarino, 1. I-56122 Pisa. ITALY.}

\cortext[cor1]{Corresponding author}

%
\begin{abstract}
To support the growing demand for data-intensive and low-latency IoT applications, Multi-Access Edge Computing (MEC) is emerging as an effective edge-computing approach enabling the execution of delay-sensitive processing tasks close to end-users. However, most of the existing works on resource allocation and service placement in MEC systems overlook the unique characteristics of new IoT use cases. For instance, many IoT applications require the periodic execution of computing tasks on real-time data streams that originate from devices dispersed over a wide area. Thus, users requesting IoT services are typically distant from the data producers. To fill this gap, the contribution of this work is two-fold. Firstly, we propose a MEC-compliant architectural solution to support the operation of multiple IoT service providers over a common MEC platform deployment, which enables the steering and shaping of IoT data transport within the platform. Secondly, we model the problem of service placement and data management in the proposed MEC-based solution taking into account the dependencies at the data level between IoT services and sensing resources. Our model also considers that caches can be deployed on MEC hosts, to allow the sharing of the same data between different IoT services with overlapping geographical scope, and provides support for IoT services with heterogeneous QoS requirements, such as different frequencies of periodic task execution. Due to the complexity of the optimisation problem, a heuristic algorithm is proposed using linear relaxation and rounding techniques. Extensive simulation results demonstrate the efficiency of the proposed approach, especially when traffic demands generated by the service requests are not uniform. 
\end{abstract}

%

\end{frontmatter}


%
\section{Introduction\label{sec:intro}}
\noindent
Recent years have witnessed the explosion of the Internet of Things (IoT) in terms not only of the number of deployed devices but also of the type of devices embedded with high-data-rate sensors (e.g., wearable devices, smartphones, vehicles)~\cite{2019_comst_city}. This new generation of IoT devices is enabling many new IoT applications, such as audio recognition, augmented reality, or cooperative driving, which involve delay-sensitive and computing-intensive processing tasks of a non-trivial amount of raw data~\cite{2019_infocom_service_data_intensive}. Besides, the widespread deployment of IoT devices has fostered the development of IoT platforms and products to support a variety of IoT-based services, such as device and network management, application development and service provisioning~\cite{2016_cs_ssn_survey}.  

In past years, the conventional approach followed by IoT platforms that offer solutions to build and deploy distributed IoT applications was data-centric and based on cloud technologies~\cite{2016_fcij_iot_platform}. Specifically, in such IoT frameworks, the streams of data that originate from IoT devices are transferred to a distant cloud infrastructure, where are stored and processed and, then, offered on-demand to multiple applications and end-users as a shared pool of virtualised sensing resources. Indeed, all top cloud-computing providers have expanded their offer with proprietary IoT solutions at all cloud levels (i.e. IaaS, PaaS and SaaS), being Azure IoT Suite\footnote{\url{https://azure.microsoft.com/en-us/overview/iot/}} and AWS IoT\footnote{\url{https://aws.amazon.com/iot}} two of the most notable examples. However, the centralised management of IoT data in the cloud has proven problematic due to the overload in the access and core network and the communication latency between end-users and the cloud. To cope with these capacity and latency constraints, as well as to fulfil the QoS requirements of new IoT applications, \emph{edge computing} has emerged as an alternative computing paradigm to support IoT applications~\cite{2017_jcna_fog_survey,2018_dcan_edge_iot,2018_commag_edge_iot}. According to this paradigm, computation resources are made available at the edge of the network, in the proximity to end-devices, significantly reducing bandwidth demands and communication latency. Different architectures have been proposed to deploy edge computing platforms. Still, the Multi-access Edge Computing (MEC) technology, which allows deploying computing resources \added{at various locations of} the RAN part of 5G mobile networks~\cite{2019_etsi_mec_arch}\added{ and, more generally, at different network edge devices such as BSs and WiFi access point}, is rapidly gaining momentum to facilitate the deployment of delay-sensitive IoT applications~\cite{2018_comst_mec_iot}. Since MEC servers are much more resource-constrained than cloud data centres, it is a compelling challenge to design efficient resource management policies to maximise the number of IoT services that can be deployed while satisfying their QoS constraints. 

\added{Given the importance of MEC for future networks, there} is \added{already an} extensive literature addressing resource-allocation problems in MEC systems when the service requests are received from mobile users. Specifically optimal solutions have been proposed for computation offloading~\cite{2016_tnet_offloading,2018_jiot_offloading,2020_jcna_offloading}, service placement~\cite{2020_coms_service_placement}, and routing of user requests to MEC servers~\cite{2020_tnet_mec}. However, existing solutions do not properly take into consideration some fundamental properties of emerging delay-sensitive and data-intensive IoT applications. \added{First, }many IoT applications require the periodic execution of computing tasks on a specific type of sensed data that is collected from \emph{dispersed data sources}, \added{and not directly uploaded by the user requesting the services that is offloaded on the edge platform.} Thus, the placement of the requested IoT service in the MEC infrastructure not only impact the utilisation of available computing resources, but it also influences the routing of IoT data within the MEC network. \added{This means that} the service placement problem \added{should be formulated considering the dependencies between data flows and data processing, as well as data redundancy}. \added{Second, }existing resource allocation algorithms for MEC system mostly assume that service requested by end-users are executed independently, use separate input data, and are assigned a different set of resources. A few recent studies have introduced the concept of \emph{shareable resources} to account for services that can share the same copy of data and code (e.g., libraries for video analytics)~\cite{2018_icds_edge_share,2019_infocom_service_data_intensive}. \added{In this work, we advocate the use of \emph{data caches} to enable application information sharing and to reduce }the bandwidth demands both in the access and core networks. \added{Finally, in many cases the owners of the IoT infrastructure and the MEC infrastructure are separated}, and they do not exchange their business-related data. For instance, the MEC system may not be aware of which data should be shared between the deployed IoT services\added{, which complicates determining a suitable mapping between data sources and computing resources.}

\added{To overcome the above-mentioned shortcomings of a MEC-based IoT system,} we propose an architectural solution to enable an IoT service provider to transparently interact with a MEC infrastructure integrated in a 5G network, which offers \emph{edge IaaS capabilities}. \added{It is important to point out that} the edge IaaS model is the most popular service model for MEC platforms~\cite{2019_netsoft_mec_iaas,2017_mec_business}. In this case, the edge resources are provided as virtual execution environments (VMs or containers) of variable sizes. Besides, virtualised and independent logical networks (called \emph{network slices}) on the same physical network infrastructure are supported. Our approach enables the steering and shaping of IoT traffic within the MEC system, and the dynamic allocation of edge resources to IoT applications. The proposed framework is compliant to the ETSI MEC architecture as it relies on an entity, that we call \emph{IoT Service Manager}, which is instantiated as a standard MEC application with specific support for data sharing between IoT services. In addition, our approach does not require the IoT service provider to disclose information to the 5G network operator about the traffic characteristics of the deployed IoT applications. 

\added{In the proposed MEC-enabled framework for IoT applications, a key problem consists in determining the MEC host that should execute the IoT services in order} to fulfil the QoS requirements of the deployed services while minimising the consumption of network and edge resources. \added{In this work, we propose a general (nonlinear) formulation of the optimal service placement, resource allocation and data management} to maximise the number of admitted services while minimising the total cost of communication, computation and storage resources that the IoT service provider shall lease from the 5G network operator. Our formulation caters for the sharing of IoT data between IoT services with overlapping geographical scope, \added{aiming} to reduce the IoT traffic data that needs to be routed between MEC servers. Our approach allows the IoT service provider to keep full control of the cost of service deployment. Moreover, we consider the heterogeneous QoS requirements of IoT applications, such as the frequency of periodic task execution, to \added{dynamically adjust the reporting rate of active data sources}. 

\added{To address the computation efficiency issues of the nonlinear problem,} we leverage linear relaxation and rounding techniques~\cite{2011_book_combinatorial} to develop a heuristic algorithm that can determine a feasible solution of the original problem. Using simulation results, we show that in many practical scenarios, our algorithm finds close-to-optimal solutions with a significantly lower time complexity than the exact solver. \added{To assess the heuristic performance}, we carry out an extensive evaluation \added{using three benchmarks, namely two greedy-based heuristics and a solution inspired by an operator placement algorithm for data stream processing applications that is proposed in~\cite{2019_TPDS_ODP_best}}. Our results show that the proposed heuristic is  more efficient that the considered benchmarks when traffic demands are not uniform. Furthermore, the proposed approach ensures a more uniform utilisation of the edge and network resources 

The rest of this paper is organised as follows. Section~\ref{sec:use_case} discusses the reference scenario and the use cases. Section~\ref{sec:mec_iot} presents the proposed MEC architectural solution. The optimisation model is developed in Section~\ref{sec:problem}. Section~\ref{sec:approximation} explains the proposed heuristic algorithm. Section~\ref{sec:results} presents the evaluation of our proposed algorithms, while Section~\ref{sec:related} reviews our contribution compared to related works.

%
\section{Architecture and Use Cases\label{sec:use_case}}
\begin{figure}[tbhp]
    \centering
    \includegraphics[trim={1cm 1cm 5cm 5cm},clip,angle=0,width=1.0\columnwidth]{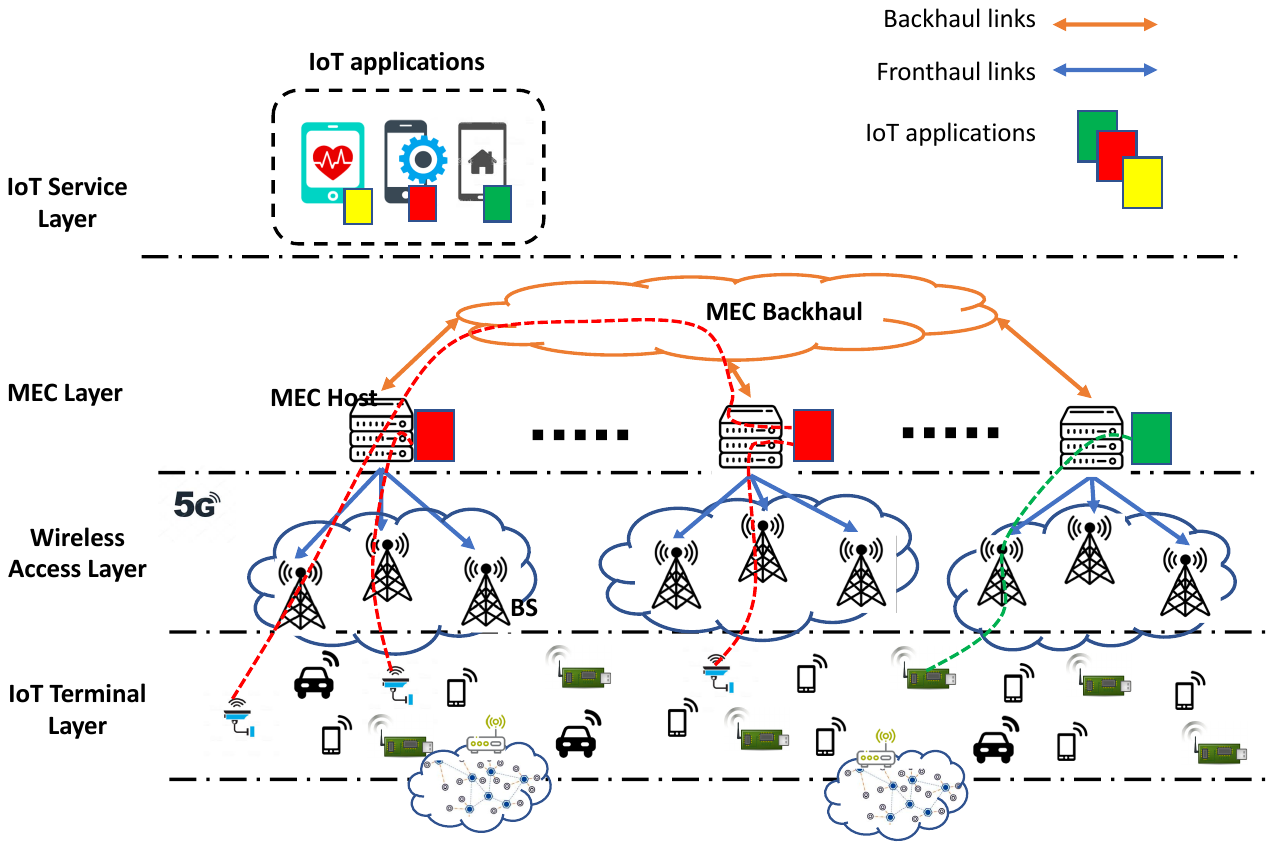}
    \caption{Illustration of the high-level system architecture. Different colours represent different types of IoT services. Dashed coloured lines represent IoT data traffic. The figure shows that three services belonging to two service types are deployed in the MEC system. }
    \label{fig:arch}
\end{figure}

\noindent
In this work, we envision a scenario in which the IoT platform offering IoT services to end-users and the 5G mobile network operator are two separate stakeholders in the IoT value chain. In the following, we describe the high-level architecture of the IoT system we consider, which is also illustrated in Figure~\ref{fig:arch}. Then, we present a simple scenario and a few application use cases motivating the considered architecture. This discussion will provide the background to explain how the IoT platform and the 5G system can transparently interact through the use of MEC technologies without exchanging critical business-related information about their customers or infrastructure deployment. 
%
%
\subsection{High-level architecture\label{sec:arch}}
\noindent
The \emph{IoT Service Provider} (\emph{IoTSP} for brevity) is an organisation (e.g. a private company, a municipality, a utility provider, etc.) that owns and manages a sensing infrastructure, which comprises a potentially large number of different \emph{IoT terminals}. We use the concept of IoT terminal as a generalisation of the classical model of IoT devices -- that is to say, things that are equipped with micro-controllers, sensing/actuating capabilities, transceivers for digital communications, and protocol stacks to connect (typically wirelessly) to the Internet. For instance, an IoT terminal could be: $i)$ a high-data-rate sensor (e.g. a surveillance camera deployed in a city), $ii)$ a multi-sensing platform (e.g., a weather station)), $iii)$ a crowdsourced mobile device (e.g. a smartphone, a connected car), or $iv)$ an \emph{IoT gateway} that acts as a proxy of the sensing resources available in an in-situ WSN deployed in the area. Then, the IoTSP may offer to end-users a \emph{catalogue of different data-intensive and delay-sensitive IoT applications} to choose from. The IoTSP's customers subscribe to a service from the pool of available ones selecting the demanded QoS parameters for the service, such as data granularity, frequency of service updates, geographical scope of the service execution~\cite{2019_mcomstd_mec_iot}. The IoTSP aims at the maximisation of its revenues by satisfying the largest number of customers' service requests. However, the IoTSP may not have the capabilities that are required to transfer, store, and process the huge amount of data which needs to be collected from geographically distributed IoT terminals to fulfil the QoS requirements of requested services. In the envisioned architecture, the IoTSP relies on the communication and edge-based computing resources that are provided by a \emph{5G Mobile Network Operator} (\emph{5G-MNO} for brevity). 

From a high-level perspective, the 5G-MNO implements a radio access network, which consists of dense deployment of heterogeneous base stations with different coverage ranges, supporting multiple communication standards (5G NR, LTE and WiFi). Then, the 5G network provides the connectivity platform enabling the IoTSP to collect the data generated from the IoT terminals under the coverage area of deployed base stations. In addition, a MEC system is integrated within the 5G network to offer edge-computing resources to its users. In this study, we assume that the \emph{5G-MNO is a provider of edge IaaS capabilities}, i.e. it provides to its customers distributed compute and storage capabilities, usually in the form of a virtualised pool of resources~\cite{2019_netsoft_mec_iaas}. Then, the IoTSP is one of the customers of the MEC system: $i)$ it specifies the resources it needs across the MEC infrastructure to deploy the IoT applications requested by end-users (e.g. virtual machines or containers based on the virtualisation technology used by the 5G-MNO), and $ii)$ it pays for resources according to use over time~\cite{2017_mec_business}. We recall that the core technical challenge we address in this study is to define a policy allowing the IoTSP to select the optimal set of network, computation and storage resources in the edge IaaS enabled by MEC platform. \added{It is worth pointing out that we have considered a 5G-MNO only as an exemplification of a mobile communication technology with specific support for IoT environments. However, the same concept could be applied to other radio access technologies and network architectures by leveraging the ability of MEC technologies to enable heterogeneous networks (see Section~\ref{sec:mec_arch} for a more detailed discussion on this aspect).} 

While the details of the MEC reference architecture will be discussed in Section~\ref{sec:mec_arch}, here we remind that one of the key components in a MEC system is the MEC host. Basically, a MEC host is a server that can be physically deployed in different locations between the base stations, a network aggregation point and the core network~\cite{2018_etsi_mec_5g}. As shown in Figure~\ref{fig:arch}, each MEC host can directly serve a given geographical area that comprises multiple base stations. Then, dedicated communication links are established between the MEC host and the base stations in its serving area (fronthaul links), between  MEC hosts (backhaul links), and between MEC hosts and the core network. The placement of the IoT application in the MEC infrastructure and the topology of the MEC network affect the routing of the traffic from the data sources to the targeted IoT applications. 
%
%
\subsection{Motivating scenario: IoT services for a Smart City\label{sec:sc}}
\noindent
Inspired by previous studies~\cite{2107_globecom_slaas,2018_comst_mec_iot,2019_mcomstd_mec_iot}, we show how a Smart City could benefit from the envisioned IoT system, and motivate the deployment of IoTSPs. Nowadays, it is a well-established concept that the IoT paradigm is a key enabler for realising the Smart City vision~\cite{2014_jiot_sc}. Indeed, an \emph{urban IoT} system is required to collect a large variety of data on city resources and infrastructures, which can then be used for the optimisation of the public services that are offered to citizens. As observed before, the advancements on IoT technologies and devices is paving the way to a variety of new business models for smart cities that require the development of IoT applications with high QoS requirements. First, there is a plethora of IoT platforms\footnote{\url{https://www.iot-survey.com/platform-provider}}
that enable almost autonomous IoT device and network management. Thus, it is increasingly common that municipalities own their urban sensing infrastructures, acting as an IoTSP for different end-users such as police, public transportation companies, domestic energy providers, and business associations. In addition, besides classical sensors for environmental monitoring, car counting, on-street parking detection, etc., the use of high-resolution video cameras, microphones, and other high-data-rate sensors have been rapidly expanding in the urban environment, opening the way to new IoT applications that involve various types of perception-related tasks~\cite{2019_comst_city}. An important example of such computation-intensive tasks that require continuous processing of massive amount of data from high data-rate sensors is audio-visual event classification. Specifically, there is a vast literature on the design of various methodologies for scene analysis and event classification that leverages audio and visual information~\cite{2007_tmm_audio_video,2019_tmc_event_recognition}. Recent advancements in ML/AI techniques \added{allow} to develop complex real-time video analysis, which can be used to identify and classify persons, objects, events, define and use event-based rules (entering/exiting area, leaving/removing objects), and counting objects (vehicles) or people. The video analytics can be further enhanced by sound data. It is important to note that different applications may have different requirements in terms of resolution of the recorded images or quality of the audio recordings, frequency of data acquisition, and spatial granularity of the multimedia data. Clearly, the higher the data quality, the higher the granularity, and the higher the data rate of the sensory data streams to be processed. In addition, enhanced video/sound analytics with real-time processing creates the need for executing the analytics functions close to the sources of these multimedia data streams for better performance and significant savings in communication latencies~\cite{2016_mce_mec_iot}.
%

%
\section{Solutions for MEC-Assisted IoT Service Providers\label{sec:mec_iot}}
\noindent
In this section, we first briefly overview the MEC specification, with a focus on the components and capabilities that are the most relevant to our solution. Then, we provide a detailed description of the proposed IoT Service Manager component, which is the MEC-based entity implementing the orchestration of IoT applications.  
\begin{figure}[tbhp]
    \centering
    \includegraphics[trim={1cm 2cm 3cm 4,5cm},clip,angle=0,width=1.0\columnwidth]{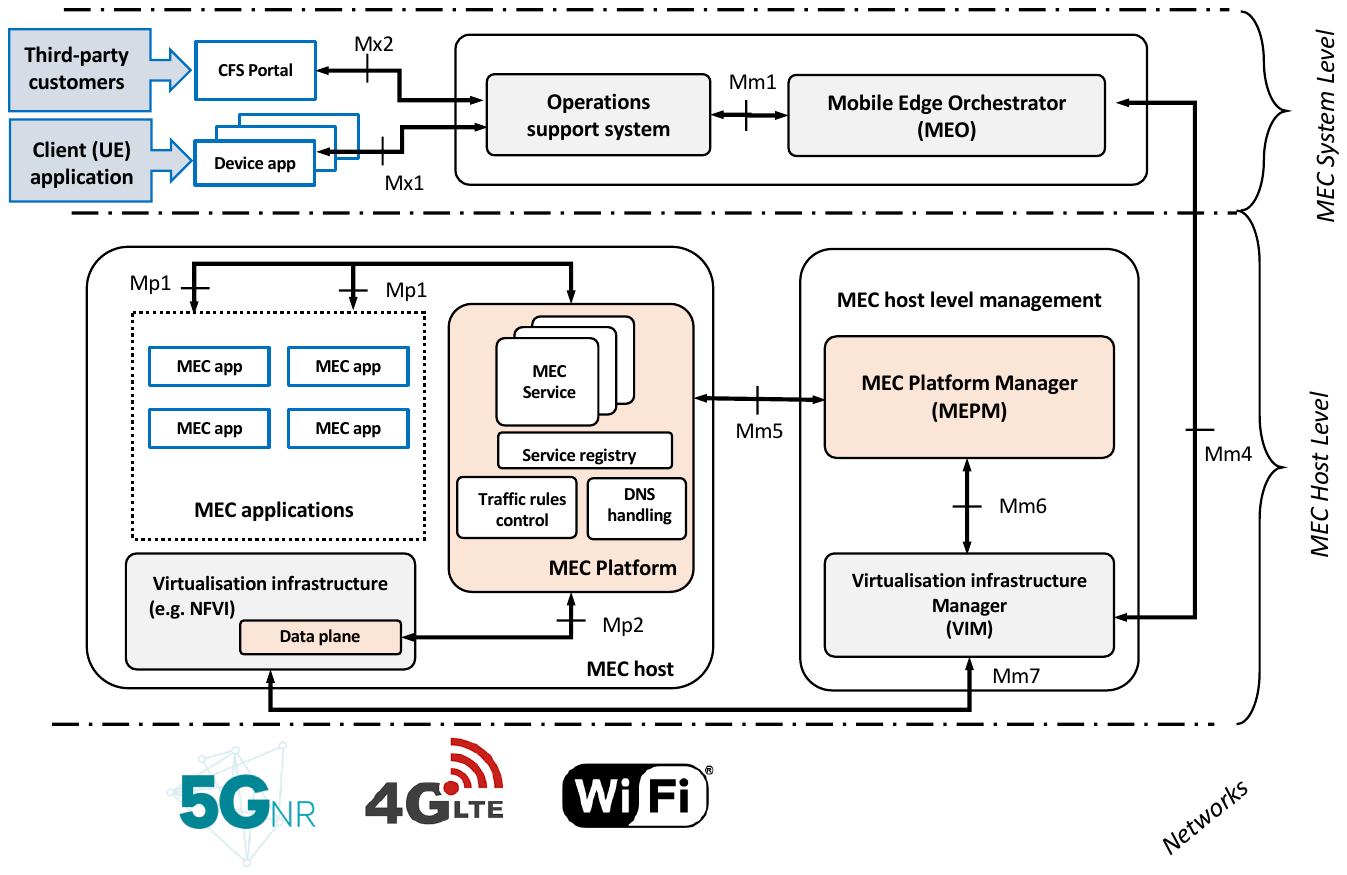}
    \caption{(\emph{Simplified}) MEC reference architecture and functional components.}
    \label{fig:mec}
\end{figure}
%
%
\subsection{ETSI MEC reference architecture. \label{sec:mec_arch}}
\noindent
\added{In 2014, ETSI established an Industry Specification Group (ISG) to specify a new standard and reference architecture, called Mobile Edge Computing (MEC), to complement the cloud radio access networks (C-RAN) of cellular operators with cloud-computing capabilities. However, over the years the scope of the MEC concept has been expanded to allow the deployment of MEC servers over multi-tier heterogeneous networks (for short, HetNets) and different elements of the network edge, such as BSs (aka eNB in 4G and gNB in 5G), optical network units, radio network controller sites, and WiFi access points~\cite{2020_access_mec_survey}. To reflect this change of perspective in the MEC concept, the ETSI MEC initiative was been renamed as Multi-access Edge Computing. Thus, not only MEC technologies will be important for 5G systems, but also for HetNets in general.} Figure~\ref{fig:mec} shows a simplified functional view of this reference architecture, focusing on interfaces and components that are the most relevant to our proposal~\cite{2019_etsi_mec_arch}. Three main logical entities constitute the core part of the MEC architecture: $(i)$ the MEC host, $(ii)$ the MEC system-level management, and $(iii)$ the MEC host-level management. Basically, a MEC host is an edge node that provides physical computing, storage and communication resources, and includes a MEC platform and a virtualisation infrastructure. The former component provides the functionalities needed to run the MEC applications on top of the virtualisation environment offered by the MEC host (e.g. through VM or container-based technologies).

Furthermore, the MEC platform enables MEC applications to provide and consume MEC services. Specifically, a MEC service is exposed with a RESTFul API, which MEC applications can subscribe to if authorised~\cite{2017_etsi_mec_part1,2017_etsi_mec_part2}. A number of MEC services have been already defined in the MEC reference architecture, which are essential enablers for our IoT Service Manager. For instance, the ``Location'' service that, when available, provides authorised applications information about the radio nodes associated to a MEC host, as well as the location of all UEs currently served by these radio nodes. The ``Bandwidth Manager'' service, when available, allows allocation of bandwidth to specific traffic routed to and from MEC applications. 

The virtualisation infrastructure of a MEC host includes a data plane that executes the traffic steering and filtering rules that are set by the MEC platform. Such traffic rules offer the capability to route and shape traffic between a specific UE and a MEC application, or among MEC applications. It is worth pointing out that the MEC platform can receive specific traffic rules from the MEC platform manager (MEPM) but also \emph{authorised MEC applications can request the activation of MEC application traffic rules dynamically}~\cite{2019_etsi_mec_arch}. As explained later, this capability is essential to allow the IoTSP to route IoT traffic between the IoT service Managers. In addition to managing traffic rules, the MEPM: $(i)$ manages the full life cycle of the MEC applications running on MEC hosts; $(ii)$ informs the MEC system management level of application-related events; and $(iii)$ provides all the basic functionalities to run applications on a specific MEC host and to enable these applications to discover, produce or consume MEC services. The MEPM operates in cooperation with the Virtualisation infrastructure manager (VIM), which allocates the virtualised resources of the MEC hosts, maintain status information on the available resources, and monitors application faults and performance. 

The core component of the control and management plane at the system level is the MEC orchestrator (MEO), which is responsible for selecting the MEC hosts to which MEC applications will be deployed based on constraints, such as latency, available resources, and available MEC services. To perform this task, the MEO maintains a broad set of information about the resource configuration and usage in the deployed MEC hosts and the topology of the entire MEC system. The service requests processed by the MEO are only the ones that are authorised by the Operating Support System (OSS) of the telco operator. The requests to run an application in the MEC system can originate from a device application, i.e. a MEC-aware component of the client application running on the user's device (e.g. UE). Alternatively, \emph{an operator's third-party customer} (e.g. an enterprise) \emph{can register with the MEC system to utilise and manage edge capabilities}. In our scenario, an IoTSP can be seen as one of these third-parties that are allowed to instantiate applications on the MEC system, paying a fee based on resource usage~\cite{2017_mec_business}. 
%
%
\subsection{A MEC-enabled IoT Service Manager\label{sec:iot_sm}}
\noindent
One of the goals pursued by an IoTSP is to maintain the control of the IoT data that is shared and routed to the IoT applications that are running in the MEC system. To achieve this objective, we envision that the IoTSP requests the 5G-MNO to host a software instance of a component, which we call \emph{IoT Service Manager} (\emph{IoT-SM} for brevity), in each MEC host of the MEC system. The primary purpose of this component is to enable the dynamic loading/\added{unloading} of IoT services. Thus, it can be regarded as a sort of \emph{middlebox} embedded into a MEC edge server specifically designed to host IoT services. However, this entity provides also several additional features: $(i)$ it dynamically \emph{scale} the usage of edge resources (memory, CPU, bandwidth) \emph{within the constraints} set by the MEC system-level management, $(ii)$ it supports the dynamic loading of IoT application services and their related data (e.g. libraries, databases, processing code); $(iii)$ it manages traffic rules to steer IoT traffic among other instances of IoT-SMs, and between local IoT services; $(iv)$ it supports a data cache and brokering service to enable reuse of IoT data between collocated IoT services; $(v)$ it shapes the traffic that is generated from the activated IoT terminals under the serving area of the MEC host; and $(vi)$ it maintains a map of the IoT terminals that are manageable from a MEC host. 

\begin{figure}[tbhp]
    \centering
    \includegraphics[trim={2cm 1,5cm 8cm 8,5cm},clip,angle=0,width=1.0\columnwidth]{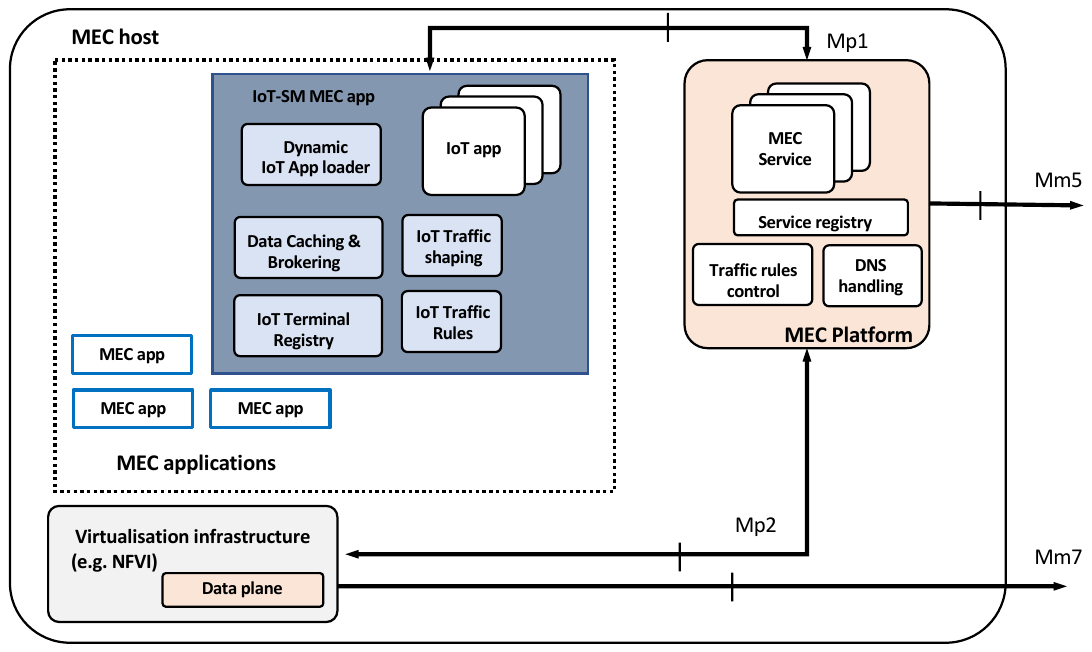}
    \caption{Functional architecture of the IoT Service Manager.}
    \label{fig:iot-sm}
\end{figure}
The functional blocks of the IoT-SM that support the above features are described in the following and illustrated in Figure~\ref{fig:iot-sm}. The \emph{IoT Terminal Registry} (\emph{RD}) functional block is in charge of maintaining a registry of the IoT terminals that are within the serving area of the MEC host on which the IoT-SM is running. \added{To build such list of available data sources, the IoT-SM relies on the above-mentioned Location Service~\cite{2017_etsi_mec_location}, which not only provides the list of UEs in a particular location area, but also the location information of all UEs currently served by the radio node(s) associated with the MEC host.} We remind that the 5G-MNO implements its specialised policy to associate the terminals to available BSs (e.g., to maximise spectral efficiency or improve load balancing~\cite{2018_tvt_cell}). Thus, the IoTSP has no control about which MEC host would serve individual IoT terminals. The \emph{RD} also communicates with its counterparts running on the other MEC hosts to build a global map of IoT terminal--MEC host associations. The \emph{Dynamic IoT App Loader} (\emph{LOADER}) functional block is the core element of the proposed framework as it handles the management of the IoT applications and their associated resources. Specifically, it checks the current status of available resources and already deployed IoT services and, if a new IoT service can be loaded without violating the constraints set by the MEO, it downloads the library/code that is needed to run the IoT application, and requests to the MEC platform the necessary additional resources. It is worth pointing out that our architectural model is amenable to both centralised and distributed formulations of the resource allocation problem. In the former case, one IoT-SM is elected by the IoTSP as the leader manager that operates as a logically centralised orchestrator. In the latter case, an overlay exists between the IoT-SMs that take local decisions on resource allocation in a distributed manner~\cite{2020_comnet_uncoordinated}. The \emph{LOADER} operates in collaboration with the \emph{Traffic Shaping} (\emph{TS}) and \emph{Data Caching \& Brokering} (\emph{DCB}) functional blocks. The former is in charge of shaping the streams of IoT data that are needed to execute the IoT applications. Specifically, it executes the decisions received by the \emph{LOADER} on which IoT terminals to activate, and which data streaming rate to assign to them. We remind that the optimal configuration of sensing resources that are assigned to an IoT application depends on the QoS parameters of the admitted applications (e.g. frequency of service execution, geographical scope of the service), and the bandwidth constraints. On the other hand, the \emph{DCB} supports the cache of IoT data to enable different IoT applications instantiated on the same IoT-SM to share and reuse the same data. Furthermore, it is responsible for route traffic between IoT applications running on different MEC hosts, thus allowing the shared use of sensing resources that can be dispersed in a large deployment area. Finally, the \emph{Traffic Rules} (\emph{TR})  functional block is in charge of informing the MEC platform of the traffic rules that are needed to redirect the traffic between the IoT-SMs. 

From the perspective of the MEC reference architecture and IoT-SM implementation, we observe that the proposed software component can be architected as a conventional MEC application, which is instantiated in the MEC system in response to a request of the IoTSP. Then, the Mp1 interface is used by the IoT-SM MEC applications to access the needed MEC services and to activate traffic rules in the MEC platform. Our architectural model is well aligned with the one specified in~\cite{2019_mcomstd_mec_iot}. However, the model given in~\cite{2019_mcomstd_mec_iot} focuses only on the virtualisation of IoT Gateways. Finally, our system solution requires that the virtualisation technology used by the MEC system to implement the virtualisation infrastructure supports dynamic resizing of VMs at run-time. This is needed to allow the running IoT-SMs to dynamically scale up and down their allocated CPU and memory resources following the downloading of new IoT applications or the termination of previously admitted IoT applications \added{, without restarting running IoT applications}. It is important to point out that vertical scaling (i.e., changing CPU and RAM resources allocated to a running server) is a feature currently offered by many cloud PaaS platforms~\cite{2020_autoscaling}.
\paragraph{Implementation considerations}\added{In past years, several experimental and open-source fog computing platforms were developed to allow IoT service providers to automatically program, manage and orchestrate dynamic data processing flows over cloud and edges, being FogFlow}\footnote{\url{https://fogflow.readthedocs.io/en/latest/}}\added{ and Stacks4Things}\footnote{\url{http://stack4things.unime.it/}} \added{two relevant examples. Following a similar development path, there are various MEC platform under development at the moment. ETSI MEC ISG has recently established a new working group, called Deployment and Ecosystem Development working group (WG DECODE), to facilitate and promote the use of open-source components to build MEC systems by allowing applications that are virtualised in the edge, to access network and users’ information from the local node using MEC-standardised open APIs}\footnote{\url{https://forge.etsi.org/rep/mec}}\added{. In addition, non-profit technology consortia around the world, such as the Open Edge Computing Initiative and the Linux Foundation, are promoting open-source projects for edge computing platforms, which also provide supporting services that can be used to develop the IoT-SM concept.}

%
\section{System Model and Problem Formulation\label{sec:problem}}
\noindent
The goal of this section is to formulate an optimisation problem for the joint service placement, allocation of resources and data management in the MEC system described above. First, we introduce the system model, the used notation (also shown in Table~\ref{tab:notation}) and the modelling assumptions. Then, we formulate the problem in terms of constraints and optimisation objectives. 
\begin{table*}[th]
\renewcommand{\arraystretch}{1.1}
\footnotesize
\begin{center}
\caption{List of notation.\label{tab:notation}}
\begin{tabular}{|C|p{0.6\textwidth}|}
\hline
Parameter & Definition \\
\hline \hline
\multicolumn{2}{|c|}{Sets, vectors} \\
\hline \hline 
$\mathcal{W}=\{c_k\}$ & set of cells; subscript $k$ refers to cell $c_k$ \\
\hline
$\mathcal{R}=\{r_i\}$ & set of sensing resources; subscript $i$ refers to resource $r_i$ \\
\hline
$\langle l(i), c(i) \rangle$ & data type measured by resource $r_i$, and cell where it is located\\
\hline
$\mathcal{O}=\{o_t\}$ & set of IoT terminals; subscript $t$ refers to terminal $o_t$  \\
\hline
$\mathcal{BS}=\{bs_h\}$ & set of base stations; subscript $h$ refers to node $bs_h$\\
\hline
$\mathcal{H}=\{h_m\}$ & set of MEC hosts; subscript $m$ refers to host $h_m$ \\
\hline
$\mathcal{BS}_m \subseteq \mathcal{BS} $ & set of base stations that are served by MEC host $h_m$ \\
\hline 
$\mathcal{S}=\{s_j\}$ & set of IoT services requested by end-users; subscript $j$ refers to service $s_j$ \\ 
\hline
$ \langle l_j,\mathcal{W}^j,\lambda_j, \delta_j, \gamma^p_j \rangle$  & Requirement vector of IoT service $s_j$: data type, geographical scope (i.e., set of cells that shall be covered to serve the $s_j$), frequency of service execution, processing load, persistent memory\\
\hline 
$\mathcal{R}_t \subseteq \mathcal{R}$ & set of sensing resources that are exposed by IoT terminal $o_t$ \\
\hline \hline
\multicolumn{2}{|c|}{Decision variables} \\
\hline \hline 
\hline 
$x_{ij}$ &  Binary variable indicating if the data collected by resource $r_i$ is delivered to the IoT service $s_j$ ($x_{ij}=1$) or not \\
\hline 
$y_{mj}$ &  Binary variable indicating if IoT service $s_j$ is deployed on the MEC host $h_m$ ($y_{mj}=1$) or not \\
\hline
$f_i$ &  sampling rate assigned to sensing resource $r_i$\\
\hline \hline
\multicolumn{2}{|c|}{Auxiliary variables, constants}  \\
\hline \hline 
$ \psi_{th}$  & binary variable indicating that terminal $o_t$ is associated to base station $bs_h$ \\
\hline 
$ \hat{b}_l$  & payload size for a sample of data type $d_l$ \\
\hline 
$F_{h}$, $F_{m}$, $F_{m_1m_2}$  & total amount of IoT data transferred to base station $bs_h$, to MEC host $h_m$, and between MEC hosts $h_{m_1}$ and $h_{m_2}$, respectively\\
\hline 
$\langle \Delta_m, \Gamma_m \rangle$  & maximum computing power and storage that can be allocated by MEC host $h_m$ to running IoT services\\
\hline 
$B^{\uparrow}_{h}$, $B_{m}$, $B_{m_1m_2}$  & uplink bandwidth capacity of base station $bs_h$, bandwidth capacity of fronthaul link to MEC host $h_m$, and bandwidth capacity of the link between MEC hosts $h_{m_1}$ and $h_{m_2}$, respectively\\
\hline 
$c^{bw}_1,c^{bw}_2,c^{cpu},c^{m}$ & unitary cost of wireless bandwidth, wired bandwidth, CPU cycles, storage \\
\hline 
\end{tabular}
\end{center}
\end{table*}
%
%
\subsection{System model\label{sec:system}}
\noindent
Our model is aligned with the architecture illustrated in Figure~\ref{fig:arch}. We consider an area $A$, where an IoTSP has deployed a set $\mathcal{O}$ of IoT terminals, which could be part of a larger in-situ sensing platform deployed in the city. An IoT terminal is described by $o_t$, $t \!=\! 1, \ldots, N$, where $N$ is the total number ot IoT terminals. Each IoT terminal $o_t$ hosts a set $\mathcal{R}_t$ of sensing resources, while $\mathcal{R} \!=\! \cup_{t \in [1,N]} \mathcal{R}_t $ is the whole set of sensing resources that are available in the reference area. Sensing resources are denoted by $r_i$. For simplicity, we assume that the area of interest is divided into a regular grid of $K$ square cells, and the sensed data collected by an IoT terminal is associated to the cell where the IoT terminal is located. Cells are denoted by $c_k$, $k \!=\! 1, \ldots, K$. Similarly to previous studies~\cite{2018_infocom_crowd_urban_sensing}, we assume that a sensing resource can only collect data of a specific data type. Let $L$ be the total number of data types of interest. Let us denote with $l(i) \!=\! 1, \ldots, L$ the data type measured by resource $r_i$, and with $c(i) \!\in\! \mathcal{W}$ the cell where the resource $r_i$ is located. Note that different attributes may be associated to a data type, such as the temporal and spatial resolution of the sensor reading, the energy dissipated by the sensor during the measurement, etc. For the sake of our analysis, we only consider the size (in bytes) of the message payload $\hat{b}_l$ that is needed to upload the sensor reading for data type $l$.

The radio access network (RAN) of the 5G-MNO comprises $H$ small base stations (SBSs) that are densely deployed to increase the network capacity and reduce latency by leveraging spatial reuse~\cite{2016_mwc_5g}. Let $\mathcal{BS} =\{bs_1, \ldots, bs_H\}$ denote the set of available SBSs. An IoT terminal can be located under the (possibly) overlapping coverage regions of multiple SBSs. It is out of the scope of this work to investigate cell association rules for base-station assignment (the interested reader is referred to~\cite{2016_comst_udn} for a comprehensive survey). We assume that the 5G-MNO associates each IoT terminal to exactly one of the nearby SBSs, and a binary variable $\psi_{t,h} \!\in\! [0,1]$ models the one-to-one mapping between IoT terminal $o_t \!\in\! \mathcal{O}$ and SBS $bs_h \!\in\! \mathcal{BS}$. 

As described in Section~\ref{sec:use_case}, the area of interest is also served by a MEC system that consists of a set $\mathcal{H}$ of MEC hosts. Each MEC host $h_m \in \mathcal{H}$ is responsible for managing a subset $\mathcal{BS}_m \subseteq \mathcal{BS}$ of the available SBSs. We assume that MEC hosts have limited storage and computing resources, and that are connected to the SBSs and among each other via links with limited capacity. Then, a portion of these resources (i.e. a \emph{slice}) is pre-allocated by the 5G-MNO to the IoTSP tenant. Specifically, each MEC host $h_m$ specifies the maximum storage capacity (i.e. hard disk) $\Gamma_m$ and the maximum computation capacity (i.e. CPU cycles) $\Delta_m$ that can be assigned to the resident IoT-SM and the collocated IoT services. In addition, the 5G-MNO assigns to the IoTSP a maximum wireless uplink bandwidth capacity $B^{\uparrow}_{h}$ to upload the IoT data from the terminals associated to SBS $b_h \!\in\! \mathcal{BS}$. The fronthaul links connecting MEC host $h_m$ to the radio nodes in $\mathcal{BS}_m$ have a limited total bandwidth equal to $B_{m}$. Finally, the backhaul link between MEC host $h_{m_1}$ and MEC host $h_{m_2}$ has fixed capacity $B_{m_1,m_2}$. In this study, we assume a \emph{flat} network topology for the MEC system, in which all MEC hosts are interconnected. \added{Investigating} hierarchical edge architectures \added{is} left \added{as} future work. For the sake of simplicity, in this work, we assume that the capacity bounds of the IoT slice are constant during the life cycle of the IoT services. In other words, while the amount of network and edge resources that are leased by the IoTSP can be dynamically adjusted, the resource limits are set by the 5G-MNO and maintained fixed.  

The IoTSP offers to its end-users a catalogue $\mathcal{S}$ of IoT services. Let $s_j$ denote the $j$-th IoT service requested by an end-user.  A requirement vector $\langle l_j,\mathcal{W}^j,\lambda_j, \delta_j, \gamma^p_j \rangle$ is associated to a service $s_j$, which describes the specific QoS parameters of the IoT service, and the resources that an IoT-SM should request to the MEC host to deploy and execute the service dynamically. Specifically, this vector defines: $(i)$ the type $l_j$ of the sensor data that is needed for providing the service, $(ii)$ the geographical scope of the service, expressed as the set $\mathcal{W}^j$ of cells that constitutes the area of interest for the demanded service, $(iii)$ the requested frequency $\lambda_j$ of periodic service execution; $(iv)$ the computation power $\delta_j$ that is needed to run the service, and ($v)$ the memory $\gamma^p_j$ that is occupied by the persistent state (i.e. code and data) of the service. 
%
%
\subsection{Problem formulation\label{sec:formulation}}
\noindent
For the sake of brevity, and when no ambiguity occurs, in the following we use the subscript index $i$ to refer to sensing resource $r_i$, the subscript index $t$ to refer to IoT terminal $o_t$, the subscript index $m$ to refer to MEC server $h_m$, the subscript index $k$ to refer to cell $c_k$, the subscript index $h$ to refer to SBS $bs_h$, and the subscript index $j$ to refer to service $s_j$. In addition, we introduce the indicator function $\mathbf{1}_{a}(X)$ of an element $a$ of a set $X$ as follows:
\begin{equation}
    \mathbf{1}_{a}(X) = \begin{cases}
    1 & a \in X \\
    0 & a \notin X \; .
    \end{cases}
\end{equation}
The IoTSP needs to decide: $(i)$ the set of sensing resources and IoT terminals that should be activated to satisfy the QoS requirements of deployed IoT services; $(ii)$ the IoT-SM in charge of loading the data and code that is necessary to execute the new IoT service; and $(iii)$ how the data collected from the activated sensing resources should be distributed to the interested IoT services over the MEC network. As explained later, an IoTSP pursues a twofold objective: the maximisation of its revenue (in terms of successfully deployed IoT services), and the minimisation of the total cost of communication, computation and storage resources leased from to the 5G-MNO. To model these decisions we introduce three sets of optimisation variables: $(i)$ $x_{ij} \in \{0,1\}$ which indicates whether the data collected from sensing resource $i$ is used as input data for the execution of service $j$ ($x_{ij} \!=\! 1$) or not; $(ii)$ $y_{mj} \in \{0,1\}$ which indicates if the IoT-SM installed in MEC host $m$ is in charge of executing service $j$ ($y_{mj} \!=\! 1$) or not; and $(iii)$ $f_i \!\in\! \mathbb R_{\ge 0}$ which specifies the frequency of data updates from resource $i$. We refer by \emph{service placement} policy to the vector:
\begin{align}
    \mathbf{y} & = \left ( y_{mj} \in \{0,1\} : m  \in \mathcal{H}, j \in \mathcal{S} \right ) \; ,
\end{align}
and \emph{data management} policy to the vectors:
\begin{align}
    \mathbf{x} & = \left (x_{ij} \in \{0,1\} : i  \in \mathcal{R}, j \in \mathcal{S} \right)  \\
    \mathbf{f} & = \left (f_i \in \mathbb R_{\ge 0} : i  \in \mathcal{R} \right) \; .
\end{align}
The service placement and data management policies need to satisfy several constraints. First, an IoT service can be loaded by the IoT-SM installed in MEC host $m$ only if this decision does not violate the capacity constraints set by the 5G-MNO. Specifically, the total computational load generated by deployed IoT services running on MEC host $m$ must not exceed the computing capacity of that host, 
\begin{align}
 \sum_{j \in \mathcal{A}} y_{mj} \times  \delta_j & \le \Delta_m  , & \forall m \in \mathcal{H} \label{cr:comp_power} \;  ,
\end{align}
and the total amount of data stored in MEC host $m$ for executing the IoT service must not exceed the memory capacity of that host:
\begin{align}
\sum_{j \in \mathcal{A}} y_{mj} \times [ \gamma^p_j + \hat{b}_{l_j} \times |\mathcal{W}_j | ]  & \le \Gamma_m ,  & \forall m \in \mathcal{H} \; . \label{cr:memory} 
\end{align}
It is worth noting that constraint~(\ref{cr:memory}) accounts for the fact that the \emph{larger the geographical scope of a service, and the greater the amount of data that needs to be collected by the IoT-SM executing the service}. Second, each IoT service must run on exactly one MEC host:
\begin{align}
 \sum_{m \in \mathcal{H}} y_{mj}  & \le 1 , & \forall j \in \mathcal{S} \; . \label{cr:app_host}  
\end{align}
Third, a sensing resource $i$ is eligible for being activated only if it covers one of the cells that are requested by a deployed IoT service: 
\begin{align}
     \sum_{\substack{ i \in \mathcal{R}\\ l(i) = l_j}}  \mathbf{1}_{c(i)}(\mathcal{W}^j) x_{ij} & = \sum_{m \in \mathcal{H}} y_{mj} & \forall j \in \mathcal{S} \label{cr:xij}  
\end{align}
A core requirement of the formulated problem is that the frequency of data updates for an activated sensing resource must be able to satisfy the QoS requirements of all IoT services sharing the data produced by that sensing resource. This condition can be expressed as follows:
\begin{align}
 f_i  & \ge  x_{ij} \times \lambda _j & \forall i \in \mathcal{R}, \forall j \in \mathcal{S} \;  \label{cr:fi_ge}   \\
 f_i  & \le  \sum_{\substack{ j \in \mathcal{S}\\ l(i)=l_j }} \mathbf{1}_{c(i)}(\mathcal{W}^j) x_{ij} \times \lambda _j  \;  &  \forall i \in \mathcal{R} \label{cr:fi_le}
\end{align}
Finally, the total bandwidth load that is generated by the IoT data traffic must not exceed the bandwidth capacity of wireless and wired connections. To express this condition we introduce three auxiliary variables: $(i)$ the total amount $F_{h}$ of IoT data uploaded to SBS $h$, $(ii)$ the total amount $F_{m}$ of IoT data uploaded to MEC host $m$ and $(iii)$ the total amount $F_{m1,m2}$ of IoT data transferred from MEC host $h_{m_1}$ to MEC host $h_{m_2}$. It holds that
\begin{align}
   F_{h} & = \sum_{t \in \mathcal{O}} \psi_{th} \left [ \sum_{i \in \mathcal{R}_t} f_i \times \hat{b}_{l(i)}\right ] \; & \forall h \in \mathcal{BS} \label{eq:fh}\\
   F_{m} & = \sum_{h \in \mathcal{BS}_m} F_{bs_h} \; & \forall m \in \mathcal{H} \label{eq:fm}
\end{align}
 \begin{align}
  F_{m_1,m_2} & = \sum_{j \in \mathcal{S}} y_{m_2j} \sum_{h \in \mathcal{BS}_{m_1}} \sum_{\substack{t \in \mathcal{O} \\ \psi_{t h} =1 }}\sum_{\substack{i \in \mathcal{R}_t \\ l(i) = l_j}} x_{ij} \lambda_j \times \hat{b}_{l(i)} & \label{eq:f12}  
\end{align}
It is worth pointing out that the set of equations~(\ref{eq:fh})-(\ref{eq:f12}) also define how the IoT traffic is routed within the MEC network. Owing to the model assumptions, it holds that the IoT traffic must not exceed the following bandwidth capacities. 
\begin{align}
 F_{h} &  \le B^{\uparrow}_h  & \forall h \in  \mathcal{BS} \label{cr:fh} \; \\
   F_{m} &  \le B_{m}  &   \forall m \in  \mathcal{H} \label{cr:fm} \; \\
  F_{m_1,m_2} &  \le B_{m_1m_2}  &   \forall m_1,m_2 \in \mathcal{H} \label{cr:f12}\;
\end{align}
The goal of the IoTSP is to maximise its net utility when satisfying the service requests of the end-users. Intuitively, there is positive revenue for each IoT application that can be successfully served. Thus, the utility of running services is given by 
\begin{equation}
    \mathbf{J}_{r} =  \sum_{j \in \mathcal{S}} \sum_{m \in \mathcal{H}} y_{mj}  \; . \label{eq:jr} 
\end{equation}
On the other hand, the \added{cost} of an IoT service is generated by the price paid by the IoTSP to the 5G-MNO for the connectivity and edge resources. The total cost of the used resource is given by
\begin{multline}
    \mathbf{J}_{edge}  =  \sum_{h \in \mathcal{BS}} c^{bw}_1  F_{h} +  \sum_{m \in \mathcal{H}} \left ( c^{cpu} \delta_m + c^{m} \vphantom{\sum_{n \in \mathcal{H} \setminus \{m\} }  F_{mn} } \gamma_m  \right . + \label{eq:jedge} \\ 
     \qquad \left . c^{bw}_2 \left ( F_{m} + \sum_{n \in \mathcal{H} \setminus \{m\} }  F_{mn} \right ) \right ) \; ,  
\end{multline}
Using the above variables and functions, we can formulate our problem as follows:
\begin{align}
\max_{\mathbf{x}, \mathbf{y}, \mathbf{f}} &  \left ( \mathbf{J}_{r} - \gamma \mathbf{J}_{edge} \right ) \label{eq:opt_problem} \\ 
\mathrm{s.t.  } & \textrm{ constraints (\ref{cr:comp_power}) -- (\ref{cr:fi_le}), (\ref{cr:fh}) -- (\ref{cr:f12})} \; , \label{eq:opt_constraints} \nonumber
\end{align}
where the weight parameter $\gamma \!\ll\! 1$ is used to assign a higher priority to the maximisation of the number of deployed services. Note that the problem~(\ref{eq:opt_problem}) is a Mixed-Integer Nonlinear Programming (MINLP) problem, which is NP-hard in general. The non-linearity of the optimisation problem is due to constraint~(\ref{cr:f12}), which involves the cross-product of decision variables. However, we can apply standard linearisation techniques to convert the problem~(\ref{eq:opt_problem}) into a Mixed Integer Linear Programming (MILP) problem~\cite{1974_OR_linearisation}. Specifically, we introduce the following set of auxiliary binary variables:  
\begin{align}
& \bm{\theta} = (\theta_{mij} \in \{0,1\}, m \in \mathcal{H}, i \in \mathcal{R}, j \in \mathcal{S}) \; . 
\end{align}
and the following linear constraints
\begin{align}
\theta_{mij} & \le x_{ij} & \forall m \in \mathcal{H}, i \in \mathcal{R}, j \in \mathcal{S} \label{cr:theta_x} \\ 
\theta_{mij} & \le y_{mj} & \forall m \in \mathcal{H}, i \in \mathcal{R}, j \in \mathcal{S} \label{cr:theta_y} \\ 
\theta_{mij} & \ge (x_{ij} + y_{mj}) - 1 & \forall m \in \mathcal{H}, i \in \mathcal{R}, j \in \mathcal{S} \label{cr:theta_xy}
%
%
\end{align}
Using the above variables, it is possible to linearise constraint~(\ref{cr:f12}) by eliminating the cross-product terms as follows:
\begin{multline}
     \sum_{j \in \mathcal{S}} \sum_{h \in \mathcal{BS}_{m_1}} \sum_{\substack{t \in \mathcal{O} \\ \psi_{t h} =1 }}\sum_{\substack{i \in \mathcal{R}_t \\ l(i) = l_j}} \theta_{m_2ij} \lambda_j \times \hat{b}_{l(i)}   \le B_{m_1m_2}   \\  
      \forall m_1,m_2 \in \mathcal{H} \label{cr:f12_new} 
\end{multline}
Owing to the new constraints, the equivalent MILP problem we can be written as:
\begin{align}
\max_{\mathbf{x}, \mathbf{y}, \mathbf{f},\bm{\theta}} &  \left ( \mathbf{J}_{r} - \gamma \mathbf{J}_{edge} \right ) \label{eq:opt_problem_MILP} \\ 
\mathrm{s.t.  } & \textrm{ constraints (\ref{cr:comp_power}) -- (\ref{cr:fi_le}), (\ref{cr:fh}) -- (\ref{cr:fm}), (\ref{cr:theta_x}) -- (\ref{cr:f12_new})} \; . \nonumber \label{eq:opt_constraints_MILP}
\end{align}
It is important to point out that the exact solution of a MILP problem can be efficiently computed using the well-established Branch-and-Bound (B\&B) algorithm~\cite{2000_book_optimisation}. Even though the worst-case complexity of such algorithms is still exponential, B\&B methods can leverage structural properties of the problem and meta-heuristics to restrict the search space, thus reducing the time needed to compute an almost-exact solution, at least for small- to medium-scale problem instances. 
%

%
\section{Heuristic Algorithm\label{sec:approximation}}
\noindent
In this section, we describe a Linear Relaxation-guided heuristic algorithm (LR for brevity) for the solution of the problem~(\ref{eq:opt_problem_MILP}). Our solution method is described in detail below and summarised in Algorithm~\ref{alg:approximation} and Algorithm~\ref{alg:sensor_alloc}. 

Our heuristic starts by solving a relaxed version of the original problem, called \emph{P1} (line~\ref{ln:lr_solveP1}). Specifically, \emph{P1} is obtained from~(\ref{eq:opt_problem_MILP}) by removing the integrality constraint of each integer decision variable with the following collection of linear constraints:
\begin{align}
    & 0 \le x_{ij} \le 1 & \forall i \in \mathcal{R}, j \in \mathcal{S} \; ,\\
    & 0 \le y_{mj} \le 1 & \forall m \in \mathcal{H}, j \in \mathcal{S}  \; ,\\
    & 0 \le \theta_{mij} \le 1 & \forall m \in \mathcal{H}, j \in \mathcal{S}, i \in \mathcal{R}  \; .
\end{align}
Let us denote with $\{x^{\dagger}_{ij}\}$, $\{y^{\dagger}_{mj}\}$, $\{\theta^{\dagger}_{mij}\}$ the optimal relaxed solution values that are obtained in polynomial time using a linear program solver. If the optimal solutions of all relaxed variables are integer values, then a valid solution is found, and the algorithm is stopped (line~\ref{ln:stop_cond}). Otherwise, we round these values to obtain an integer solution, denoted by $\{\widehat{x}_{ij}\}$, $\{\widehat{y}_{mj}\}$\footnote{We remind the $\theta_{mij}$ is an auxiliary decision variable used to remove the non-linear constraints with cross-product of variables. Thus, $\{\widehat{x}_{ij}\}$ and $\{\widehat{y}_{mj}\}$ do not depend on $\{\theta^{\dagger}_{mij}\}$.}. To do so, we start from an empty solution set (line~\ref{ln:init_empty_sol_var}), and then tentatively try to allocate new services. It is important to point out that we leverage the fractional solution of \emph{P1} to express the probability that rounding a variable yields a resource allocation that is feasible and optimal. In addition, the algorithm keeps track of the utilisation of network and edge resources (line~\ref{ln:init_res_var}) to discard assignments that would violate capacity constraints.

\emph{The algorithm uses the fractional value $y^{\dagger}_{mj}$ to define the order with which the services should be deployed}. Specifically, let $j^{\prime}$ denote the service with the highest probability $y^{\dagger}_{mj}$ to be deployed on a host $m$, for which an accept/reject decision is still pending (line~\ref{ln:check_best_j}). Then, the algorithm iteratively checks whether there is \added{a host} having a probability $y^{\dagger}_{mj^{\prime}} > \mu $ (line~\ref{ln:check_mj}) on which the service can be deployed without violating the problem constraints (line~\ref{ln:cr_notviolated_1}). The algorithm uses $\mu$ as a \emph{cut-off} threshold to filter out unlikely assignments, thus reducing the number of iterations that are needed to find a sub-optimal solution. The order of host allocation follows the probability $y^{\dagger}_{mj^{\prime}}$. If a feasible allocation on host $m^{\prime}$ exists, then $\widehat{y}_{m^{\prime}j^{\prime}}$ is rounded to one (line~\ref{ln:round_y1_1}). Afterwards, the algorithm checks if a feasible allocation of sensing resources that covers all the cells in $\mathcal{W}^{j^{\prime}}$ exists. If yes, the algorithm updates accordingly the utilisation of edge resources (line~\ref{ln:update_edge}). On the contrary, if there is not a feasible allocation of sensing resources, $\widehat{y}_{m^{\prime}j^{\prime}}$ is rounded back to zero. 
\algnewcommand{\algorithmicgoto}{\textbf{go to}}
\algnewcommand{\Goto}[1]{\algorithmicgoto~\ref{#1}}
\begin{algorithm}[thbp]
\renewcommand{\algorithmicrequire}{\textbf{Input:}}
\renewcommand{\algorithmicensure}{\textbf{Output:}}
\small
  \caption{- Heuristic Algorithm using linear relaxation and rounding techniques.}\label{alg:approximation}
  \begin{algorithmic}[1]
  \Require \emph{P1} \Comment{linear relaxation of original problem~(\ref{eq:opt_problem_MILP})} 
  \Ensure $\{\widehat{x}_{ij}\},\{\widehat{y}_{mj}\},\{\widehat{f}_{i}\}$  \Comment{integer assignments and sampling rates}

\Statex
\State solve \emph{P1} to obtain optimal solution $\{x^{\dagger}_{ij}\}$, $\{y^{\dagger}_{mj}\}$, $\{\theta^{\dagger}_{mij}\}$ \label{ln:lr_solveP1}
\If {$\left (x^{\dagger}_{ij},y^{\dagger}_{mj},\theta^{\dagger}_{mij} \in  \mathbb{Z} \right )$} \label{ln:stop_cond}
    \State $\widehat{x}_{ij} \gets x^{\dagger}_{ij}, \widehat{y}_{mj} \gets y^{\dagger}_{mj}$ ;
    \State \textbf{return} 
\EndIf
\Statex
\State $\Omega \gets  \{\emptyset\}$ \Comment{Set of tested services}
\State $\widehat{x}_{ij} \gets 0,\widehat{y}_{mj} \gets 0,\widehat{f}_{i} \gets 0$ \label{ln:init_empty_sol_var} \Comment{Initialise output variables}
\State $\widehat{C}_m,\widehat{S}_m,\widehat{F}_h,\widehat{F}_m,\widehat{F}_{m_1m_2}$ \label{ln:init_res_var} \Comment{Auxiliary variables used to track resource utilisation}
\Statex
\While{$(\mathcal{S} \setminus \Omega \ne \emptyset)$} \label{ln:loop_j}
    \State $ j^{\prime}\ \gets \argmax{m \in \mathcal{H}, j \in \mathcal{S} \setminus \Omega} (y^{\dagger}_{mj})$ \Comment{higher priority is given to service $j$ with the largest $y^{\dagger}_{mj}$} \label{ln:check_best_j}
    \State $\Sigma \gets  \{\emptyset\}$ \Comment{Set of tested hosts}
    \State $hostAllocated \gets \texttt{False}$
    \ForAll{$(m \in \mathcal{H} , y^{\dagger}_{mj^{\prime}} > \mu)$} \label{ln:check_mj}
        \State $ m^{\prime}\ \gets \argmax{m \in \mathcal{H} \setminus \Sigma} (y^{\dagger}_{mj^{\prime}})$ \Comment{higher priority is given to host $m$ with the largest $y^{\dagger}_{mj^{\prime}}$} \label{ln:check_best_mj}
        
        \State $\Sigma  \gets \Sigma  \cup \{m^{\prime}\}$
        \If{$\Call{CheckEdgeConstr}{m^{\prime},j^{\prime}} $} \label{ln:cr_notviolated_1} \Comment{check if constraints~(\ref{cr:comp_power}) and~(\ref{cr:memory}) are not violated when deploying service $j^{\prime}$ on host $m^{\prime}$}
            \State $hostAllocated \gets \texttt{True}$
            \State $\widehat{y}_{m^{\prime}j^{\prime}} \gets 1$ \label{ln:round_y1_1}
            \State \textbf{break} \Comment{exit for cycle}
        \EndIf
    \EndFor
    \Statex
    
    \If {($hostAllocated  = \texttt{True}$)} \label{ln:sensor_alloc}
        \If {($\Call{AllocateSensors}{m^{\prime},j^{\prime}}  = \texttt{True}$)} \label{ln:sensor_alloc_res}
            \State $\widehat{S}_m\gets \widehat{S}_m + \gamma_{j^{\prime}} \; ; \; \widehat{C}_m\gets \widehat{C}_m + \delta_{j^{\prime}}$ \label{ln:update_edge}
        \Else
            \State $\widehat{y}_{m^{\prime}j^{\prime}} \gets 0$ 
            \label{ln:unset_y1_1}
        \EndIf
    \EndIf
    \Statex
    \State $\Omega \gets \Omega \cup \{j^{\prime}\}$
    
\EndWhile
\end{algorithmic}
\end{algorithm}

Algorithm~\ref{alg:sensor_alloc} describes the procedure we implement to search for an allocation of sensing resources that can cover service $j^{\prime}$ if deployed on host $m^{\prime}$ without violating the bandwidth constraints. At the beginning, Algorithm~\ref{alg:sensor_alloc} initialises auxiliary variables with the current state of bandwidth utilisation (line~\ref{ln:init_temp_band_and_freq}). Then, for each cell $k$ in the set $\mathcal{W}^{j^{\prime}}$ that is not yet covered (line~\ref{ln:loop_cell}), the procedure iteratively searches for the best sensing resource (i.e. the resource $i^{\prime}$ with the highest $x^{\dagger}_{ij^{\prime}}$ value) to use for covering cell $k$. As in Algorithm~\ref{alg:approximation}, $\mu$ is a cut-off threshold used to filter out unlike assignments. Let us assume that the algorithm tentatively selects resource $i^{\prime}$. First of all, a frequency $f^{\dagger}_{i^{\prime}}$ is selected to fulfil the requirement on frequency of data updates for service $j^{\prime}$ (line~\ref{ln:tentaive_fi}). Then, the algorithm checks if this sampling rate assignment violates one of the bandwidth constraints (line~\ref{ln:cr_notviolated_2}). If not, the sensing resource $i^{\prime}$ is allocated to service $j^{\prime}$ (i.e., $\widehat{x}_{i^{\prime}j^{\prime}}$ is rounded to one), the cell is labelled as covered, the utilisation state of network links is temporarily updated, and the algorithm directly passes to process the next cell in set $\mathcal{W}^{j^{\prime}} \setminus \widehat{\mathcal{W}}$. On the contrary, if cell $k$ cannot be covered by anyone of the available sensing resources we exit the main loop (line~\ref{ln:failed_coverage_1}), all $\widehat{x}_{ij^{\prime}}$ are rounded to zero (line~\ref{ln:undo_i}), and the function notifies the failure of sensor allocation to Algorithm~\ref{alg:approximation} (line~\ref{ln:return_false}). Finally, if all cells in $\mathcal{W}^{j^{\prime}}$ are successfully covered, the function commits the updates of state variables (line~\ref{ln:commit_updates}), and it notifies Algorithm~\ref{alg:approximation} that the allocation of sensing resources to service $j^{\prime}$ is successful (line~\ref{ln:return_true}).  
\begin{algorithm}[tb]
\renewcommand{\algorithmicrequire}{\textbf{Input:}}
\renewcommand{\algorithmicensure}{\textbf{Output:}}
\small
    \caption{- Function used to find a sensor allocation to cover service $j$ without violating bandwidth constraints}\label{alg:sensor_alloc}
    \begin{algorithmic}[1]
    \Function{AllocateSensors}{$m^{\prime},j^{\prime}$}
    \Statex
    \State $F^{\dagger}_h \gets \widehat{F}_h,F^{\dagger}_m \gets \widehat{F}_m,F^{\dagger}_{m_1m_2} \gets \widehat{F}_{m_1m_2}$  \label{ln:init_temp_band_and_freq}
    
    
     \State  $\widehat{\mathcal{W}} \gets  \{\emptyset\}$ \Comment{set of tested cells}
        \While {$(\mathcal{W}^{j^{\prime}} \setminus \widehat{\mathcal{W}} \ne \emptyset)$} \label{ln:loop_cell}
            \State $\textrm{pick } k \textrm{ at random in } \mathcal{W}^{j^{\prime}} \setminus \widehat{\mathcal{W}}$ \label{ln:pick_point}
            \State $R_{k} \gets  \{\emptyset\}$
            \State $cellCovered \gets \texttt{False} $
                \ForAll{$(i \in \mathcal{R} , x^{\dagger}_{ij^{\prime}} > \mu)$}
                    \State $i^{\prime} \gets \argmax{i \in \mathcal{R} \setminus R_{k},l(i) = l_{j^{\prime}}, c(i) = c_k }(x^{\dagger}_{ij^{\prime}})$ \label{ln:check_best_sensor}
                    
                    \State $R_{k}  \gets R_{k}  \cup \{i^{\prime}\}$
                    \State $ f^{\dagger}_{i^{\prime}} \gets ( \lambda_{j^{\prime}} \le \widehat{f}_{i^{\prime}} \; ? \; \widehat{f}_{i^{\prime}}  : \lambda_{j^{\prime}} )$ \Comment{set tentative sampling rate for resource $i^{\prime}$} \label{ln:tentaive_fi}
                    
                    \Statex
                    \If{$\Call{CheckNetworkConstr}{f^{\dagger}_{i^{\prime}}}$} \label{ln:cr_notviolated_2}
                        \State $cellCovered \gets \texttt{True}$ \label{ln:cell_covered}
                        \State $\widehat{x}_{i^{\prime}j^{\prime}} \gets 1$ \label{ln:round_xi}
                        \State $ [F^{\dagger}_h,F^{\dagger}_m,,F^{\dagger}_{m_1m_2}] \gets \Call{IncBand}{\widehat{x}_{ij},\widehat{y}_{mj},f^{\dagger}_{i^{\prime}}}$ \label{ln:update_temp_band}
                        \State \textbf{break} \Comment{exit for cycle} \label{ln:goto_next_cell}
                    \Else
                        \State $f^{\dagger}_{i^{\prime}} \gets \widehat{f}_{i^{\prime}}$ \Comment{revert to state before line~\ref{ln:tentaive_fi}} \label{ln:undo_i_sens_freq}
                    \EndIf
                \EndFor
            
            \If {$cellCovered = \texttt{False}$} 
                \State \textbf{break} \Comment{exit while cycle} \label{ln:failed_coverage_1}
            \EndIf
            \State $\widehat{\mathcal{W}} \gets \widehat{\mathcal{W}} \cup \{k\}$
        \EndWhile
    
    \If {$cellCovered = \texttt{False}$}
        \State $\widehat{x}_{ij^{\prime}} \gets 0 \quad \forall i \in \mathcal{R}$ \Comment{Undo allocation} \label{ln:undo_i}
        \State \textbf{return \texttt{False}} \label{ln:return_false} 
    \Else
        \State $\widehat{F}_h \gets F^{\dagger}_h,\widehat{F}_m \gets F^{\dagger}_m , \widehat{F}_{m_1m_2} \gets F^{\dagger}_{m_1m_2}$ 
        \State $\widehat{f_i} \gets f^{\dagger}_i$ \Comment{Commit updates} \label{ln:commit_updates} 
        \State \textbf{return \texttt{True}} \label{ln:return_true} 
    \EndIf
    \EndFunction
\end{algorithmic}
\end{algorithm}

\added{
The worst-case time complexity, say $T(\emph{A1})$, of algorithm \ref{alg:approximation} can be expressed as follows
\begin{equation}
   T(\emph{A1}) = T(\emph{P1}) + \mathcal{O}(|\mathcal{S}| \times (|\mathcal{H}| + T(A2)) \; , 
\end{equation}
where $T(A2)=(|\mathcal{W}| \times |\mathcal{R}|)$ is the worst-case execution time of Algorithm \ref{alg:sensor_alloc}. Specifically, the first term $T($\emph{P1}$)$ is the time required to solve the linear relation of the original problem~(\ref{eq:opt_problem_MILP}), called \emph{P1}, which is polynomial in the number of variables and constraints. The second (polynomial) term accounts for the time needed to build the integral solution from the relaxed one. In the worst case, we have to check that the constraints are not violated on every  MEC host ($\mathcal{|H|}$) and from all the sensing resources in every cell ($|\mathcal{W}| \times |\mathcal{R}|$ in Algorithm \ref{alg:sensor_alloc}). Then, this test should be repeated for each service, i.e. $|\mathcal{S}|$ times. It is worth pointing out that, in practise, the average time complexity of the proposed heuristic, can be much lower as the \emph{cut-off} threshold allows us to reduce the number of service-host and cell-sensor pairs that have to be tested, trading-off accuracy with speed. Furthermore, each service may require to cover only a subset of the available cells, which also reduces the average execution time of the heuristic. }

%
\section{Results and Discussion\label{sec:results}}
\begin{figure}[th]
    \centering
    \includegraphics[trim={1cm 4cm 0cm 4cm},clip,angle=-90,width=0.7\columnwidth]{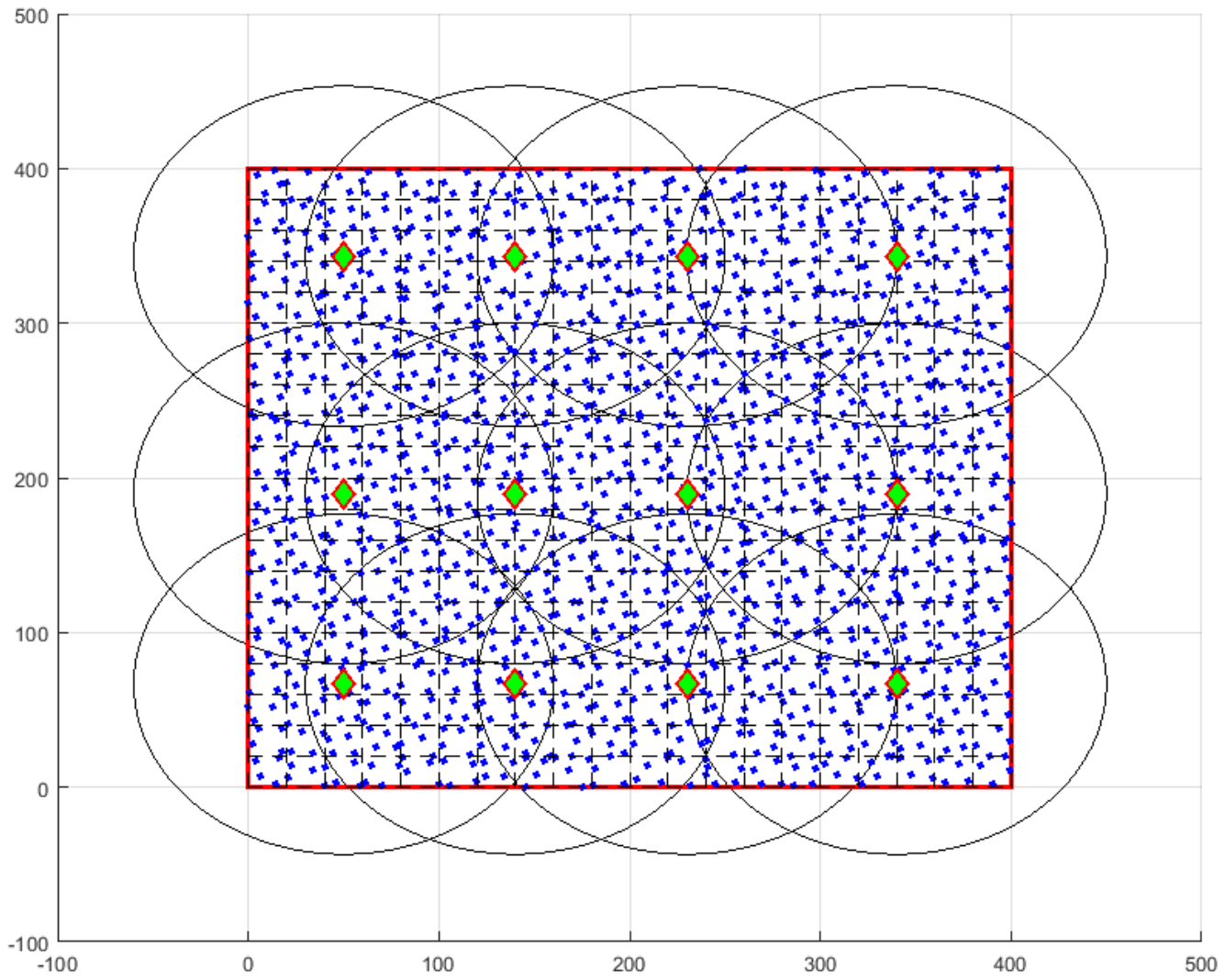}
    \caption{Evaluation setup}
    \label{fig:setup}
\end{figure}

\noindent
In this section, we evaluate the performance of the proposed solution. \added{In this study, we resort to simulation as it offers a controllable and repeatable environment. The custom-based simulation environment of the system described in Section~\ref{sec:problem} is developed using Matlab. The model in~\ref{sec:formulation} is also solved using Matlab.} We adopt an evaluation setup similar to~\cite{2020_tnet_mec}, which is also depicted in Figure~\ref{fig:setup}.  More precisely, we consider a reference area of size $400m \!\times\! 400m $, which is divided into a regular grid of $K \!=\! 400$~cells with side 20~meters. Within this area, $H \!=\! 12$~SBSs are also regularly deployed in a grid-like topology. Each SBS has a coverage radius of 120~meters. $N \!= 1200$~IoT terminals are distributed uniformly at random over the reference area. IoT terminals are associated with the nearest SBS. Without loss of generality, we assume that a MEC host is collocated with each SBS, i.e., $|\mathcal{BS}_m| \!=\! 1$. We define two base configurations for the \emph{resource slice} assigned by the 5G-MNO to the IoTSP tenant, which are summarised in Table~\ref{tab:slice}. Specifically, \emph{Scen A} represents a use case with abundant resources. For each MEC host $m$, we set the memory capacity to $\Gamma_m \!=\! 20$GB, the computation capacity to $\Delta_m \!=\! 4$GHz, the uplink wireless capacity $B^{\uparrow}_h \!=\! 60$Mbps, and the backhaul link capacity $B_{m_1,m_2} \!=\! 10$Mbps. On the other hand, \emph{Scen B} represents a use case with more constrained resources. Basically, we halve the bandwidth and computation resources, and we reduce the memory capacity to $\Gamma_m \!=\! 15$GB. We point out that a backhaul link can be implemented as a virtual link/tunnel in the metro core network of the 5G-MNO. As described in Section~\ref{sec:system}, in this study, we consider a flat network topology for the MEC system. Thus, the virtual mesh network interconnecting the MEC hosts would require a total bandwidth of $H(H-1) \!\times\! B_{m_1,m_2}$. Note that in the following evaluation, the values of the base resource configurations are also varied to assess the impact of each constraint on the system performance.    
\begin{table}[th]
    \centering
    \footnotesize
    \renewcommand{\arraystretch}{1.3}
    \begin{tabular}{| l | c | c  |}
    \hline 
    & \emph{Scen A} & \emph{Scen B} \\
    \hline
    \hline
    $\Gamma_m$ & 20~GB & 15~GB \\
    \hline
    $\Delta_m$ & 4~GHz & 2~GHz \\
    \hline
    $B^{\uparrow}_h$ & 60~Mbps & 30~Mbps  \\
    \hline
    $B_{m_1,m_2}$ & 10~Mbps & 5~Mbps \\
    \hline
    \end{tabular}
    \caption{Base configurations for resource slices assigned by the 5G-MNO to the IoTSP tenant. }
    \label{tab:slice}
\end{table}
As reference scenario, we focus on data-intensive and latency-sensitive IoT applications that require the sensing, delivery and processing of \emph{large volumes of video and sound data}~\cite{2020_tnet_mec, 2018_jiot_ssn}. To this end, we assume that each IoT terminal is equipped with high-quality microphones and cameras. Then, we consider two types of IoT services, namely visual scene recognition (VR), and audio event classification (AC). The first one requires the acquisition of high-resolution video frames that are compared with a database of possibly thousands of images. We assume that the size of a video frame is 2Mb and that end-users request video frames to be processed with a frequency from 0.125 to 1~fps. This design choice corresponds to bandwidth for video frame uploading from 0.25 to 2~Mbps. The computation is set in the range $[0.001,0.08]$~GHz per cell, assuming 40 CPU cycles per bit, and up to 4~GB of storage are consumed. The audio event classification (AC) service is similar to VR. In this case, acoustic models are compared to sound data collected from the on-board microphones~\cite{2020_mta_audio}. AC services are typically less resource-demanding than VR services. Specifically, we assume that the size of a sound recording is 1~Mb (half of a high-quality video frame), and that end-users request sound data to be processed with a frequency from 0.125 to 1~Hz. The computation of AC services consumes 30 CPU cycles per bit and require up to 2~GB of storage. Finally, the geographical scope of both AC and VR services is fixed to 10~cells.
\begin{table*}[th]
    \centering
    \footnotesize
    \renewcommand{\arraystretch}{1.3}
    \begin{tabular}{l l l l l l }
    \hline 
     Service $j$ & Data $\hat{b}_l$ (Mb) & Frequency $\lambda_j$ (Hz)& Computation (cycles per bit) & Storage $\gamma^{p}_j$ (GB) & $|\mathcal{W}^j|$\\
    \hline
        \emph{visual scene recognition} (\emph{VR})& 2 &  [0.125,1] & 40 & [1,4] & 10 \\
        \emph{audio event classification} (\emph{AC}) & 1 & [0.125,1]  & 30 & [1,2] & 10 \\
        \hline
    \end{tabular}
    \caption{Resource requirements of IoT services}
    \label{tab:req_service}
\end{table*}
%
%
\subsection{\added{Benchmarks}}
\noindent
\added{In the following, we present in detail the thee benchmarks, namely, Greedy First-Fit, Greedy Best-Fit, and DSP,} we use for performance comparison with our algorithm (called \emph{Opt}) and the approximation algorithm described in Section~\ref{sec:approximation} (called \emph{LR}).
%
\subsubsection{\added{Greedy First-Fit (\emph{G-FF})}}
\noindent 
\added{G-FF} iteratively selects an IoT service $j$ at random in the set $\mathcal{S}$, and for each $k \in \mathcal{W}^j$ greedily searches for a sensing resource that can cover the cell $k$ with the requested frequency of data updates. If all cells in $\mathcal{W}^j$ can be covered without violating the constraints on the wireless bandwidth of SBS, this algorithm tries to place the service on the MEC host that collects data from the largest fraction of cells requested by that service. If this assignment is not feasible (due to a lack of computing, storage or network backhaul resources), service $j$ is simply discarded.
%
%
\subsubsection{\added{Greedy Best-Fit (\emph{DSP})}}
\noindent 
\added{G-BF algorithm uses the same approach of the G-FF policy to allocate sensing resources to an IoT service $j$ selected at random in the set $\mathcal{S}$. Differently from G-FF, this algorithm selects the MEC host where to place the service from a sorted list $L$ in a best-fit manner.} \added{Specifically, the heuristic adds to a list $L$ all the available MEC hosts and sorts them in descending order of the fraction of cells in $\mathcal{W}^j$ that are covered by each MEC host. Then, the heuristic selects the MEC host from $L$} based on the host rank until computing, storage and wired bandwidth resources are available. If there is not a feasible allocation \added{in the set $L$}, service $j$ is simply discarded.
%
\subsubsection{\added{DSP}}
\noindent 
\added{The third benchmark is inspired by the optimal operator placement algorithm for data stream processing applications proposed in~\cite{2019_TPDS_ODP_best}. Specifically, the problem formulated in~\cite{2019_TPDS_ODP_best} aims to place processing elements that carry out a specific operation on a fixed set of data streams on a set of computing nodes in order to minimise overall expected response time, availability, and network usage. There is a clear analogy between a data stream and a sensing resource and between an operator and an IoT service. Specifically, the DSP algorithm sorts all the IoT services in ascending order of the overall data traffic that is requested by the service. Then, sensing resources are statically associated to IoT services using the same rules of the G-BF and G-FF schemes. For each IoT service from the sorted list the DSP algorithm solves the optimisation problem defined in~\cite{2019_TPDS_ODP_best} considering only the network usage in the objective function. If there is not a feasible allocation for the IoT service, that service is discarded and the DSP algorithm moves forward to allocate the next service in the ordered list.}
%
%
%
\subsection{Homogeneous scenario\label{sec:homogenous}}
\noindent
The first set of results focuses on investigating the efficiency of resource utilisation in a homogeneous scenario where each IoT service selects the cells of interest uniformly at random in the reference area, and end-users select randomly IoT services only from the VR class. Then, we conduct a stress test by evaluating the percentage of deployed IoT services versus the number of service requests, using the resource slice of \emph{Scen A} in Table~\ref{tab:slice}. The goal of these experiments is to highlight the trends in the utilisation of networking and edge resources when each MEC host collects data from a similar number of activated sensing resources. 
\begin{figure}[th]
    \centering
    \includegraphics[trim={1cm 5cm 1cm 4cm},clip,angle=0,width=0.95\columnwidth]{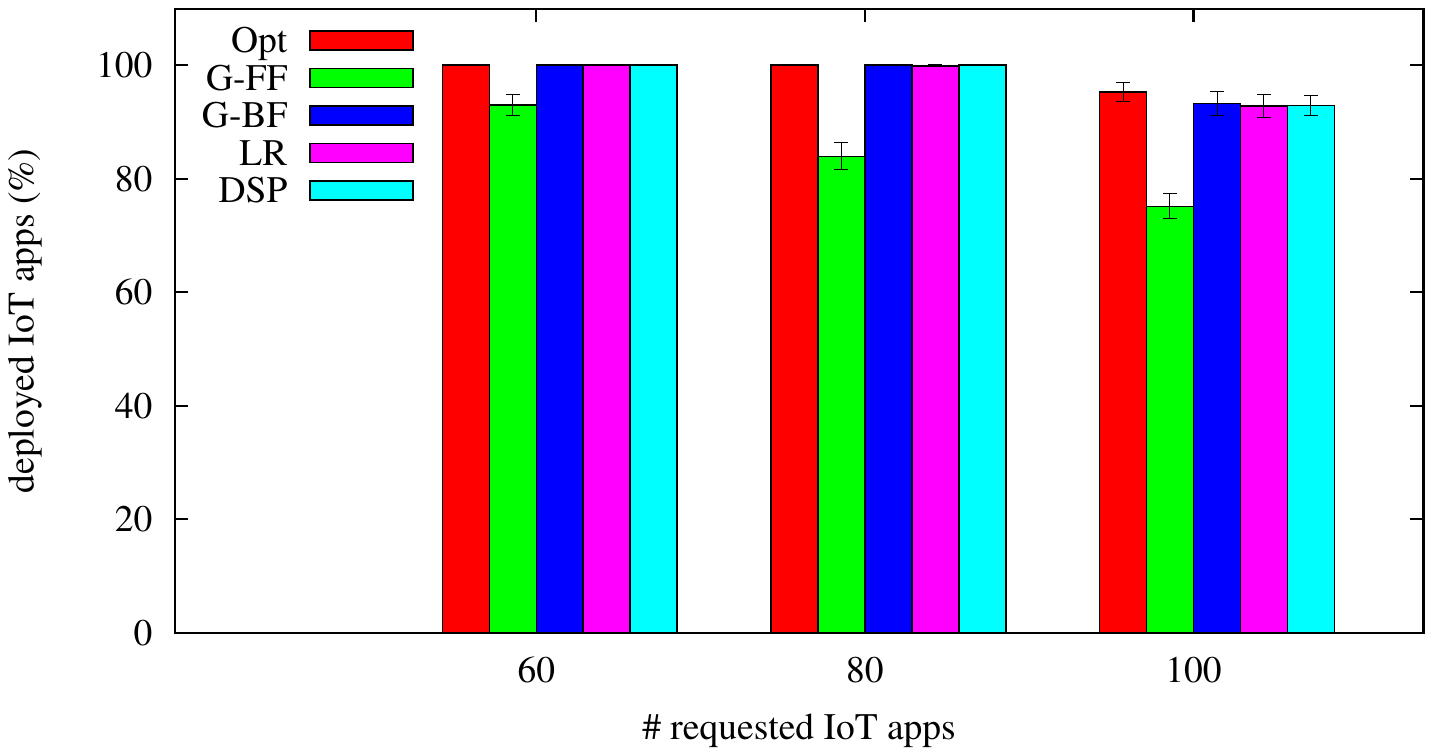}
    \caption{Percentage of deployed IoT services versus the number of service requests for uniform cell popularity (\emph{Scen A}).}
    \label{fig:app_homog}
\end{figure}

\begin{figure}[th]
\centering
  \subfloat[$|\mathcal{S}| = 60$]{
    \includegraphics[trim={1cm 7cm 0cm 7cm},clip,angle=0,width=1.0\columnwidth]{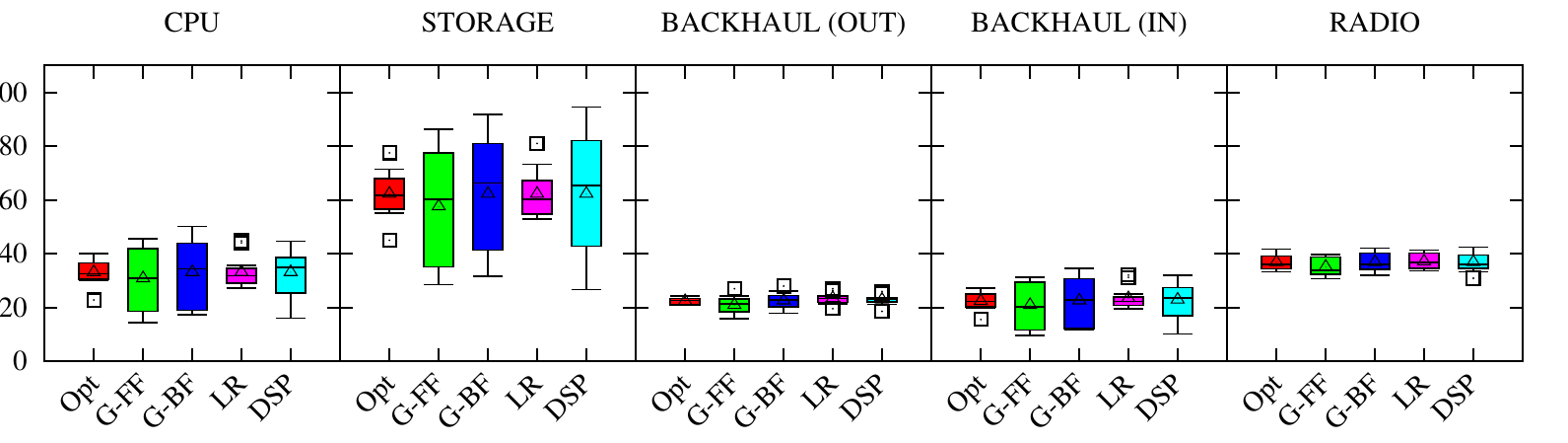}
    \label{fig:resource_homog_60}
  }
  \hfill
  \subfloat[$|\mathcal{S}| = 100$]{
    \includegraphics[trim={1cm 7cm 0cm 7cm},clip,angle=0,width=1.0\columnwidth]{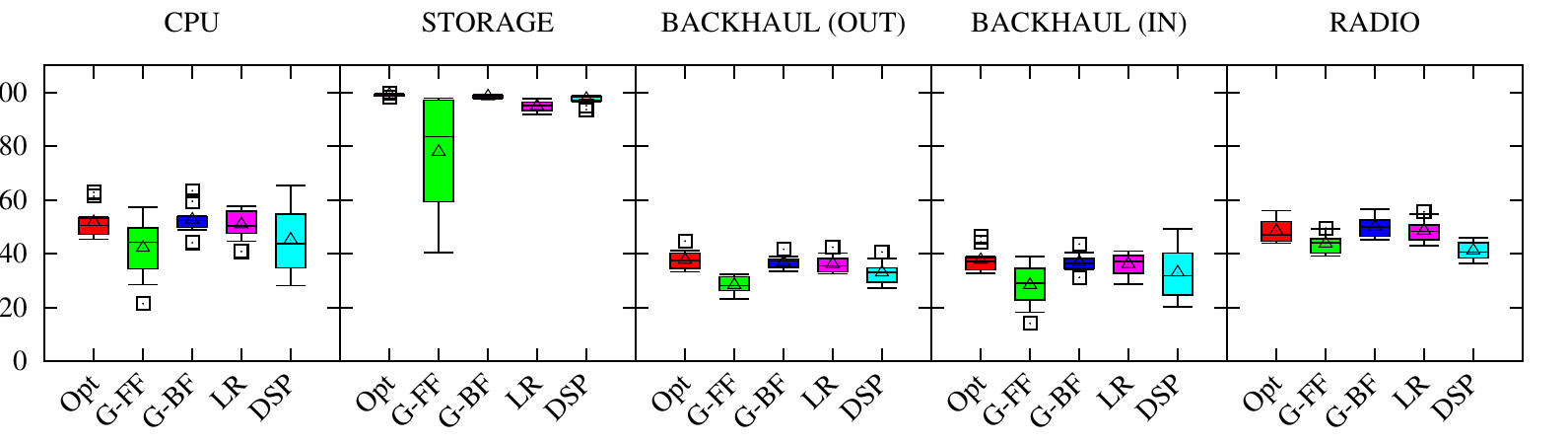}
    \label{fig:resource_homog_100}
  }
  \caption{ Boxplots of resource usage in percentage. Triangle points are the mean value. The lower and upper whiskers are the $5\%$ and $95\%$ percentile, respectively. Boxes are the outliers. \label{fig:resource_homog}}
  \vspace{-0.25cm}
\end{figure}

Figure~\ref{fig:app_homog} shows that all policies are able to admit 60 IoT services. However, the performance of \added{G-FF} rapidly degrades, and with 100 service requests, the deployment ratio is below 70\%. With 100~requests Opt performs slightly better than \added{G-BF}, LR and DSP, with a gain up to 4.3\%. To explain these results, we take a closer look into the utilisation of resources of individual MEC servers. Figure~\ref{fig:resource_homog} show the resource utilisation of each resource type using a boxplot for 60 and 100 service requests. Boxplots are a standard method to represent a statistical distribution of values in a compact way. In the graphs, we plot the uplink wireless bandwidth, the total incoming and outgoing traffic load of a MEC host, and the whole storage and computation load (as a percentage). We observe that Opt and the heuristic algorithm are able to ensure a uniform resource utilisation between MEC hosts. At the same time, the greedy-based policies \added{and DSP} experience much higher variability in resource utilisation\added{, especially as far as storage and computing power are concerned}. Furthermore, for most of the MEC hosts, the \added{benchmarks} utilise more storage and computation resources than Opt and LR. Finally, with 100 service requests all the available storage resources are utilised, and this resource bottleneck causes services requests to be discarded. The superior performance of Opt in terms of admitted applications can be explained by the more efficient service placement found by Opt. Interestingly, in the tested configurations networking resources are largely underutilised, and there are no significant differences between the different policies. Clearly, the more service requests and the larger the volume of IoT data. However, this increase is not linear with the number of service requests thanks to data caching for services with overlapping geographical scopes. 
%
%
\subsection{Impact of cell popularity\label{sec:cell_pop}}
\begin{figure}[th]
    \centering
    \includegraphics[trim={1cm 4cm 1cm 4cm},clip,angle=0,width=0.95\columnwidth]{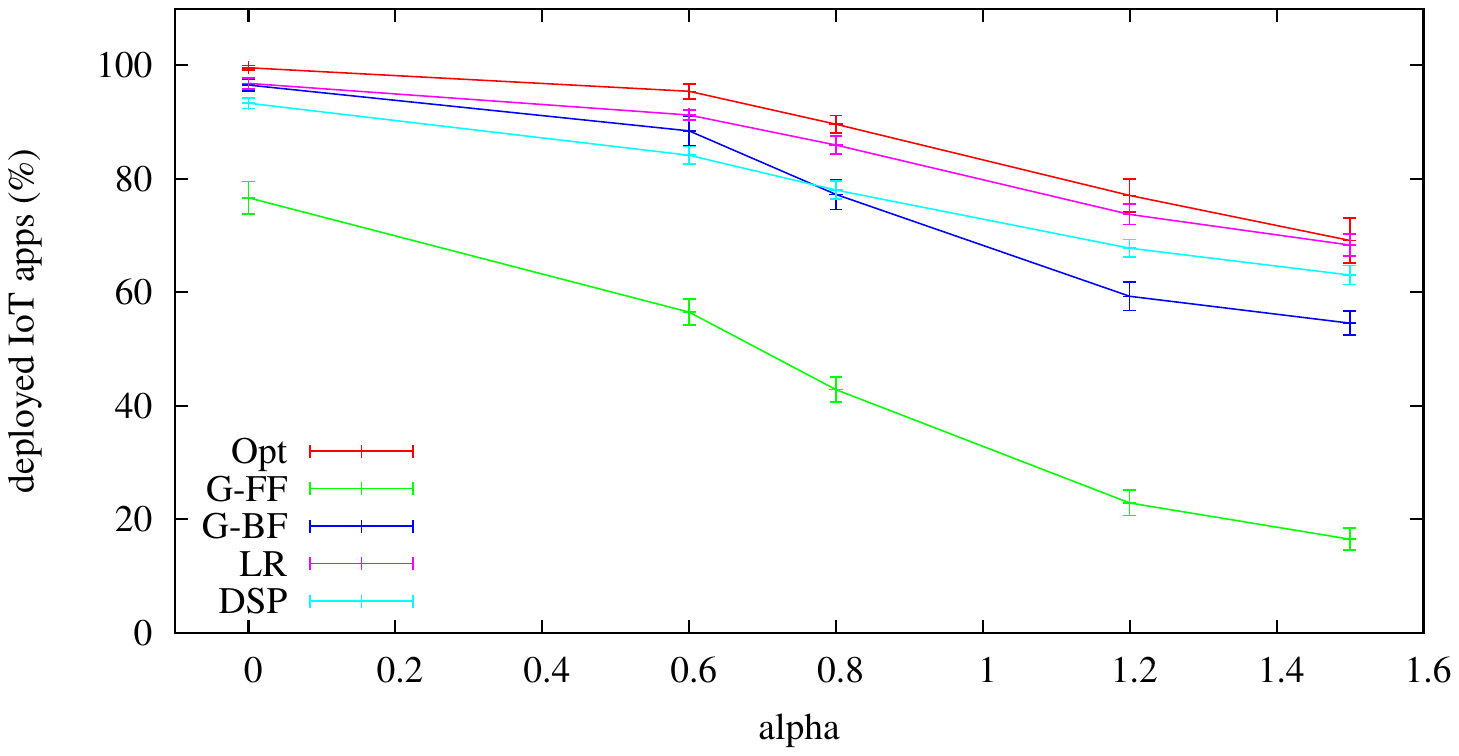}
    \caption{Percentage of deployed IoT services for different values of zipf $\alpha$ parameter. The results are obtained with two PoIs and \added{60} IoT requests (\emph{Scen B}).}
    \label{fig:app_zipf}
\end{figure}

\noindent
The main lesson we learned from the above experiments is that when the traffic demands are uniformly distributed at random over the network, and the application workloads are homogeneous, a greedy allocation strategy can still achieve near-optimal performance. However, in a real-world urban setting the states to classify or the events to detect are not uniformly distributed in the area of interest, but there are locations more popular than others (a.k.a. point of interests - PoIs). Without loss of generality, we assume that in the reference area $A$ there are two POIs and that the cell popularity around each PoI follows a Zipf distribution with shape parameter $\alpha$\footnote{When multiple PoIs exist, the cell popularity distributions generated by each POIs are simply superimposed and then re-normalised.}. The results in Figure~\ref{fig:app_zipf} are obtained using the resource slice of \emph{Scen B} in Table~\ref{tab:slice}, which ensures to have enough resources to satisfy at least 60 service requests in the homogeneous case (i.e. $\alpha = 0$) \added{using the optimal allocation policy}. 
As expected, the higher the scaling value of the zipdf distribution, the smaller the number of deployed IoT applications, with Opt that goes from a service satisfaction ratio of 100\% with $\alpha=0$ (homogeneous scenario) to 72\% with $\alpha=1.5$. This is due to the fact an increasing number of applications requests data from the most popular cells, and there may not be enough bandwidth capacity to route the IoT traffic in the MEC network towards IoT applications that are deployed on farther MEC hosts, where computation and storage resources are still available. However, Opt performs significantly better than the greedy-based solutions and \added{DSP}, with gains up to 400\% over \added{G-FF}, up to 33\% over \added{G-BF} and \added{up to 20\% over DSP}. We can observe that the performance gap between Opt and the greedy-based solutions increases with $\alpha$, \added{while the gap between Opt and DSP remains quite stable}. \added{It is important to point out that the optimal policy explores the full solutions space and determine the largest set of applications that can be admitted. On the contrary, the other schemes tries to allocate the IoT services one by one following a fixed order. Specifically, DSP sorts the IoT services based on the amount of data requested by each service, while LR leverages the optimal relaxed solutions of the original problem. The results confirm the effectiveness of our approach and the quality of the solutions computed by the proposed approximation algorithm being the performance gap between Opt and LR lower than 7\%. Interestingly, we can observe that DSP performs better than G-BF when increasing the $\alpha$ value, i.e, increasing the popularity of cells around the PoIs. Two reasons can be identified to explain this behaviour. First, when $\alpha$ is large, a service requires data from sources that are spatially concentrated and the backhaul links to/from the MEC hosts that are close to the PoIs rapidly saturate, becoming a resource bottleneck. Thus, DSP ensures a more efficient utilisation of network resources as it formulates the operator placement problem in such a way to minimise the network usage. Second, DSP gives priority to services with low data demands, which is beneficial to admit more services.}

\begin{figure}[th]
\centering
  \subfloat[$\alpha = 0.6$]{
    \includegraphics[trim={0cm 7cm 0cm 7cm},clip,angle=0,width=1.0 \columnwidth]{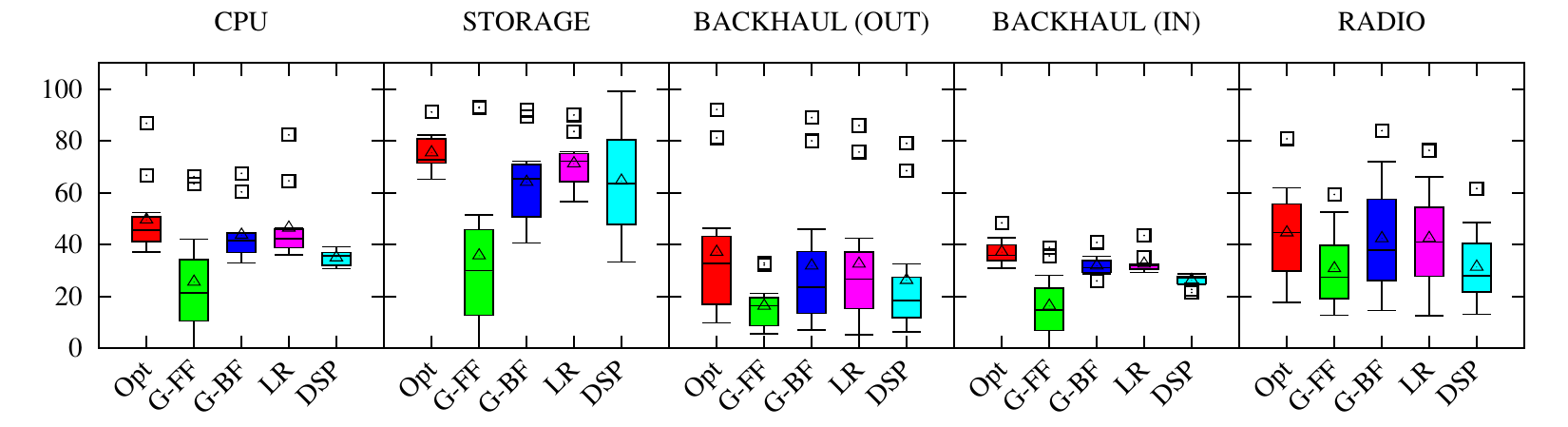}
    \label{fig:resource_zpif_06}
  }
  \hfill
  \subfloat[$\alpha = 1.5$]{
    \includegraphics[trim={0cm 7cm 0cm 7cm},clip,angle=0,width=1.0\columnwidth]{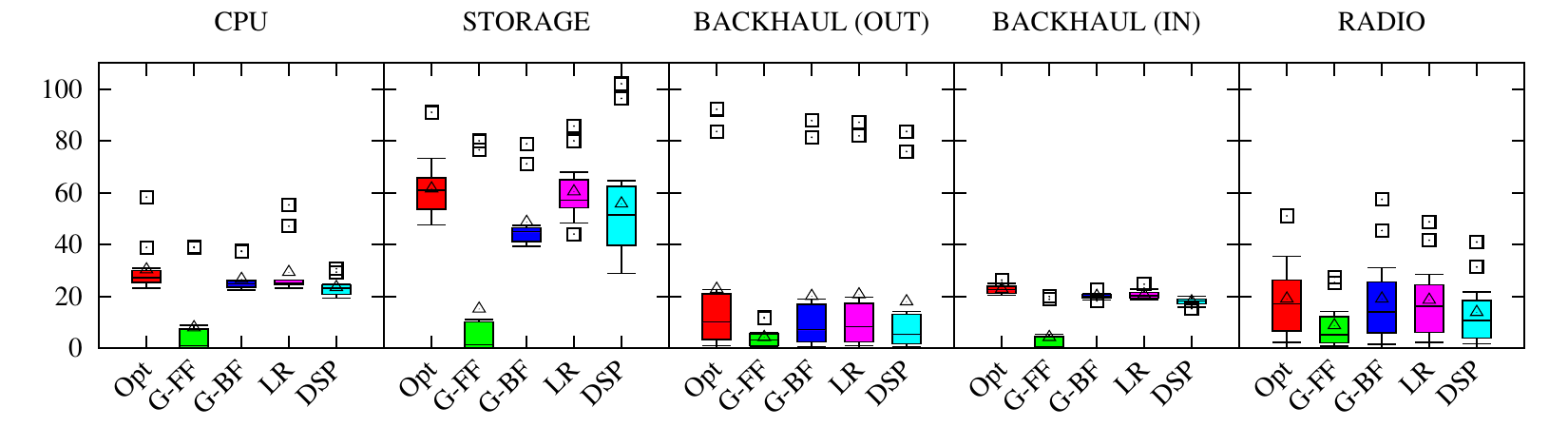}
    \label{fig:resource_zpif_15}
  }
  \caption{Boxplots of resource usage in percentage. Square points are the mean value. The lower and upper whiskers are the $5\%$ and $95\%$ percentile, respectively. Boxes are the outliers. \label{fig:resource_zpif}}
  \vspace{-0.25cm}
\end{figure}

Figure~\ref{fig:resource_zpif} shows the utilisation of each resource type for $\alpha =0.6$ and $\alpha = 1.5$ using a boxplot. Differently from the homogeneous scenario, we can observe that an increase in the popularity concentration, the average usage of computation, storage and bandwidth resources decreases. This is due to the fact that a lower number of IoT services is admitted and fewer resources are needed to execute them. Clearly, this decrease is remarkable with \added{G-FF}, as the service satisfaction ratio is very small at $\alpha = 1.5$. Interestingly, although Opt admits the largest number of IoT services, the utilisation of backhaul and radio bandwidth for most MEC hosts is similar to the one observed with \added{G-BF}. \added{Finally, DSP shows a higher variability in the usage of storage resources for $\alpha =0.6$, since this policy gives priority to services that process a low amount of data.} To explain the observed behaviours we can note that the optimal data management policy of Opt allows activating the sensing resources that can maximise the caching opportunities for admitted applications. Furthermore, the optimal service placement allows reducing the volume of traffic that needs to be delivered between MEC servers. An important difference with respect to the homogeneous case is that there are MEC hosts that are significantly more utilised than others. These overloaded MEC hosts are typically the ones closer to the PoIs, where most of the sensory streams originate. \added{The benchmarks} aim to locate the services as close as possible to the PoIs. On the contrary, Opt allows a more uniform use of the available resources, which not only delays the occurrence of resource bottlenecks but also allows to take full advantage of the available resources.   

\begin{figure}[th]
\centering
  \subfloat[$\alpha = 0.6$]{
    \includegraphics[trim={4cm 2cm 4cm 2cm},clip,angle=0,width=0.5\columnwidth]{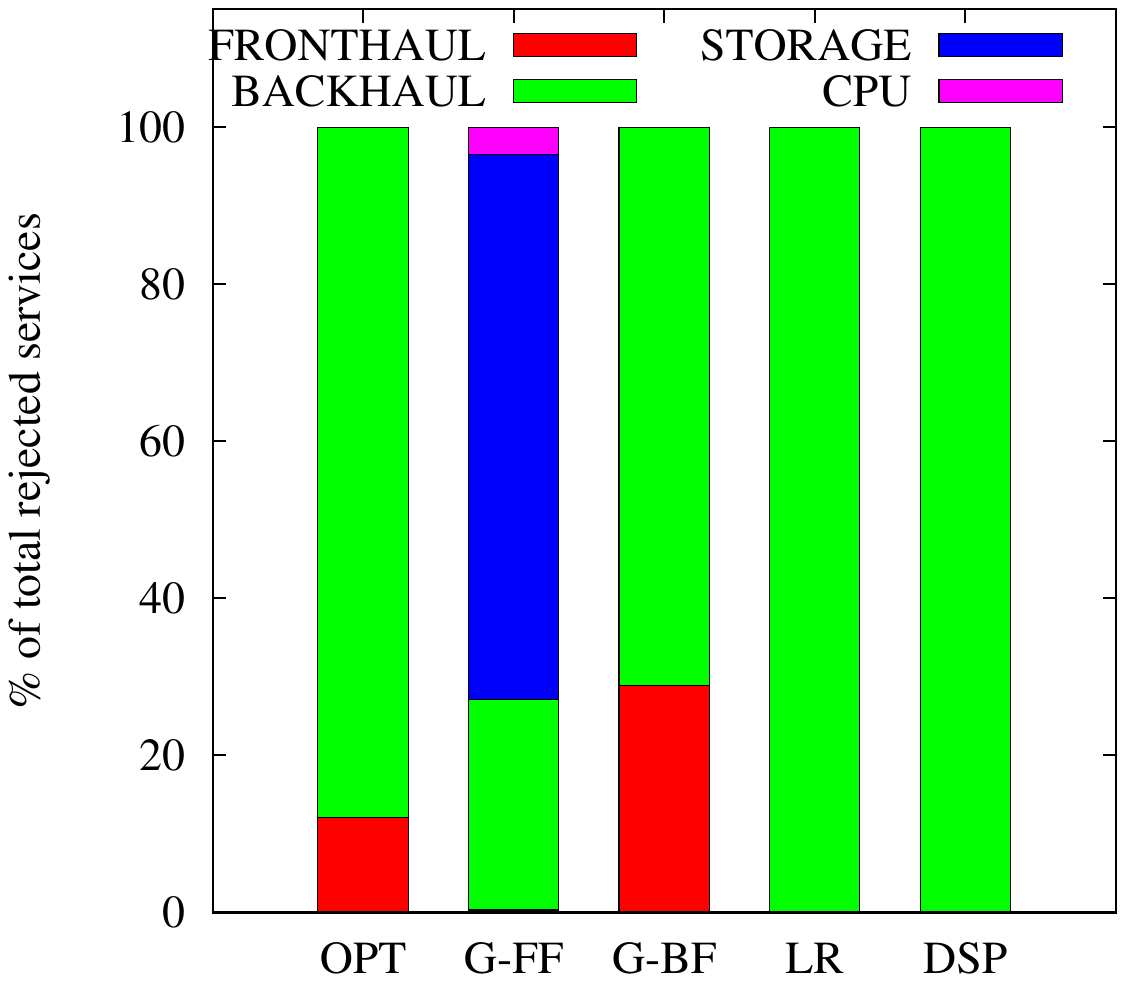}
    \label{fig:cause_zpif_06}
  }
  \subfloat[$\alpha = 1.5$]{
    \includegraphics[trim={4cm 2cm 4cm 2cm},clip,angle=0,width=0.5\columnwidth]{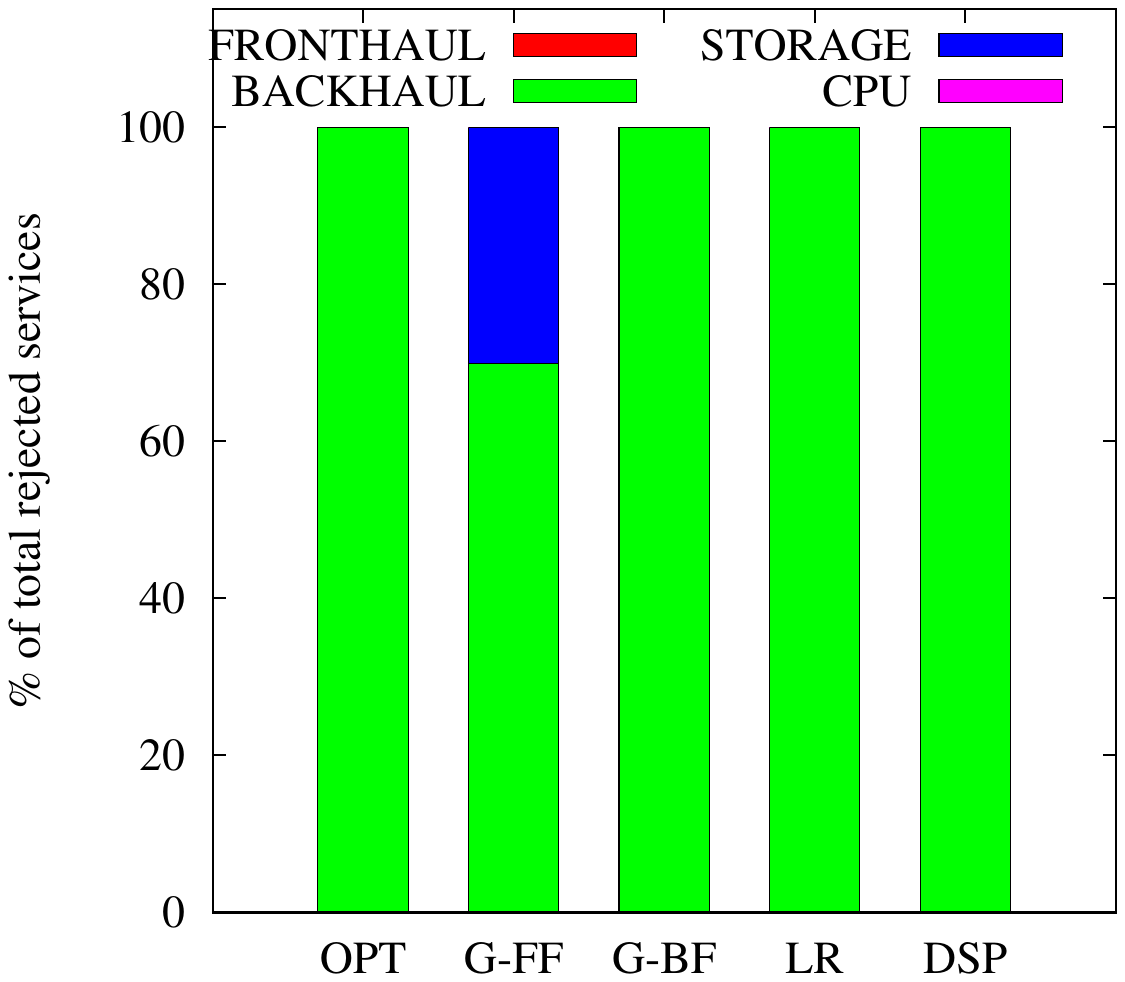}
    \label{fig:cause_zpif_15}
  }
  \caption{Distribution of the main cause of service reject for different cell popularity values. \label{fig:cause_zpif}}
  \vspace{-0.25cm}
\end{figure}

To investigate more in-depth the efficiency of the proposed solutions, Figure~\ref{fig:cause_zpif} depicts the distribution of the leading causes of service reject for different values of the zipf $\alpha$ parameter. More precisely, we consider each rejected service request, and we conduct the following tests: $(i)$ if there is not any MEC host with enough compute (memory) resources to accommodate the new service, we assume that CPU (storage) constraints are the main cause of service reject, respectively; $(ii)$ if the service can be deployed in at least one MEC host, but not all the requested cells can be covered (due to lack of sensing resources or not enough wireless bandwidth to collect the data updates from activates sensing resources) we assume that the fronthaul constraint is the main cause of service reject; finally $(iii)$ if the service can be deployed in at least one MEC host, and all the requested data can be collected from the activated sensing resources, but there is not enough wired bandwidth to deliver it to the MEC host where the service is deployed, we assume that the banckhaul constraint is the main cause of service reject. As shown in Figure~\ref{fig:cause_zpif_06} and Figure~\ref{fig:cause_zpif_15}, \added{G-FF} is the only allocation policy that rejects service requests due to unavailability of CPU and storage constraints. On the contrary, when cell popularity is highly concentrated around the PoIs, the backhaul limitation is the only cause of service reject for all the other policies. These results confirm our claim that considering the dependencies at the data level when placing services in the MEC system is essential to minimise the flow of data exchanged effectively. 

We finally highlight the running times of the presented algorithms, summarised in Table II. These running times are based on a Matlab implementation running on an HP ProDesk desktop with 3.6 GHz Intel Core i7700 processor and 16GB RAM. The exact solution of the optimisation problem is computationally demanding as a branch-and-bound method is used. In addition, the time complexity rapidly increases with the $\alpha$ parameter as more backhaul links need to be tested. For instance, the running time when $\alpha$ goes from zero (homogeneous scenario) to 0.6 increases by 480\%. When $\alpha=1.5$, all the iterations of the exact solver reach the imposed time limit (12 hours).  As expected, the greedy-based allocation algorithms require significantly lower time to return a solution. However, the proposed heuristic requires only slightly more time than a greedy-based approach to return a close-to-optimal solution and is only up to 4.3 times slower than the \added{G-BF} policy in the worst case. \added{The running times of DSP are comparable with those of LR, although the computational efficiency of DSP improves as the $\alpha$ value increases. }It is worth pointing that the running times of LR  are compatible with an online execution of the allocation policy, while Opt is suitable only for offline allocation. 
\begin{table}[th]
\footnotesize
\renewcommand{\arraystretch}{1.2}
    \centering
    \begin{tabular}{l c c c}
    \hline
    {Algorithm} & \multicolumn{3}{c}{{Running time (seconds)}} \\
     & $\alpha = 0 $ & $\alpha = 0.6 $ & $\alpha = 1.5 $ \\
     \hline
         OPT &  $722.52 \; (\pm 345)$ & $ 3470 \; (\pm 1300)$ & $43200 \; (\pm 0)$\\
         \added{G-FF} &  $12.92 \; (\pm 0.89)$ & $8.15 \; (\pm 0.39)$ & $3.64 \; (\pm 0.38)$\\
         \added{G-BF} &  $13.09 \; (\pm 1.9)$ & $8 \; (\pm 0.63)$ & $3.3 \; (\pm 0.3)$\\
         LR &  $12.77 \; (\pm 0.69)$ & $13.3 \; (\pm 0.78)$ & $14.15 \; (\pm 1.8)$ \\
        \added{DSP} &  $13.81 \; (\pm 2.2)$ & $12.53 \; (\pm 0.6)$ & $8.6 \; (\pm 0.3)$ \\
         \hline
    \end{tabular}
    \caption{Running times for different algorithms. Table reports mean and variance (in parenthesis) values measured over the 25 runs.}
    \label{tab:running_times}
\end{table}
%
\subsection{Impact of resource constraints\label{sec:constraints}}
\begin{figure*}[th]
\centering
  \subfloat[Computation capacity]{
    \includegraphics[trim={2cm 4cm 1cm 4cm},clip,angle=0,width=0.45\columnwidth]{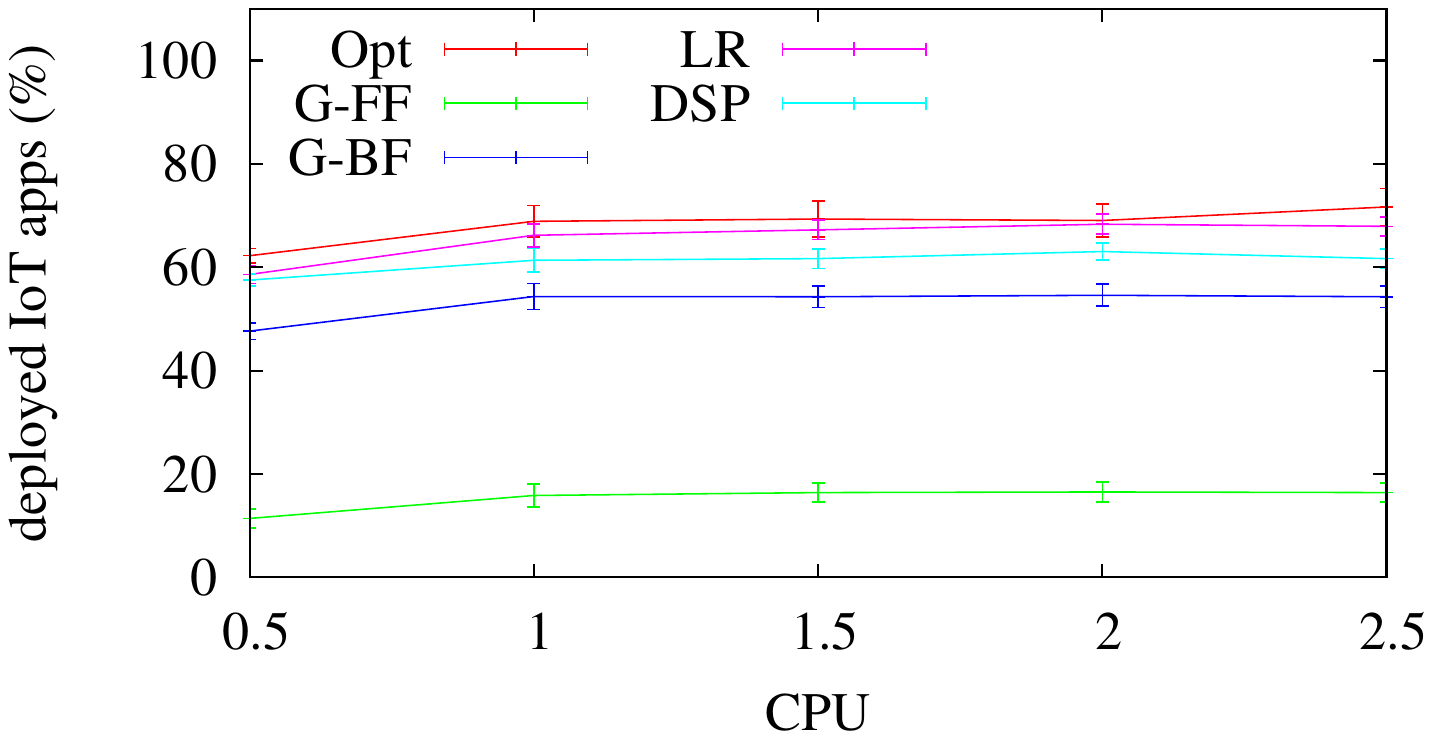}
    \label{fig:impact_cpu}
  }
 \subfloat[Storage capacity]{
    \includegraphics[trim={2cm 4cm 1cm 4cm},clip,angle=0,width=0.45\columnwidth]{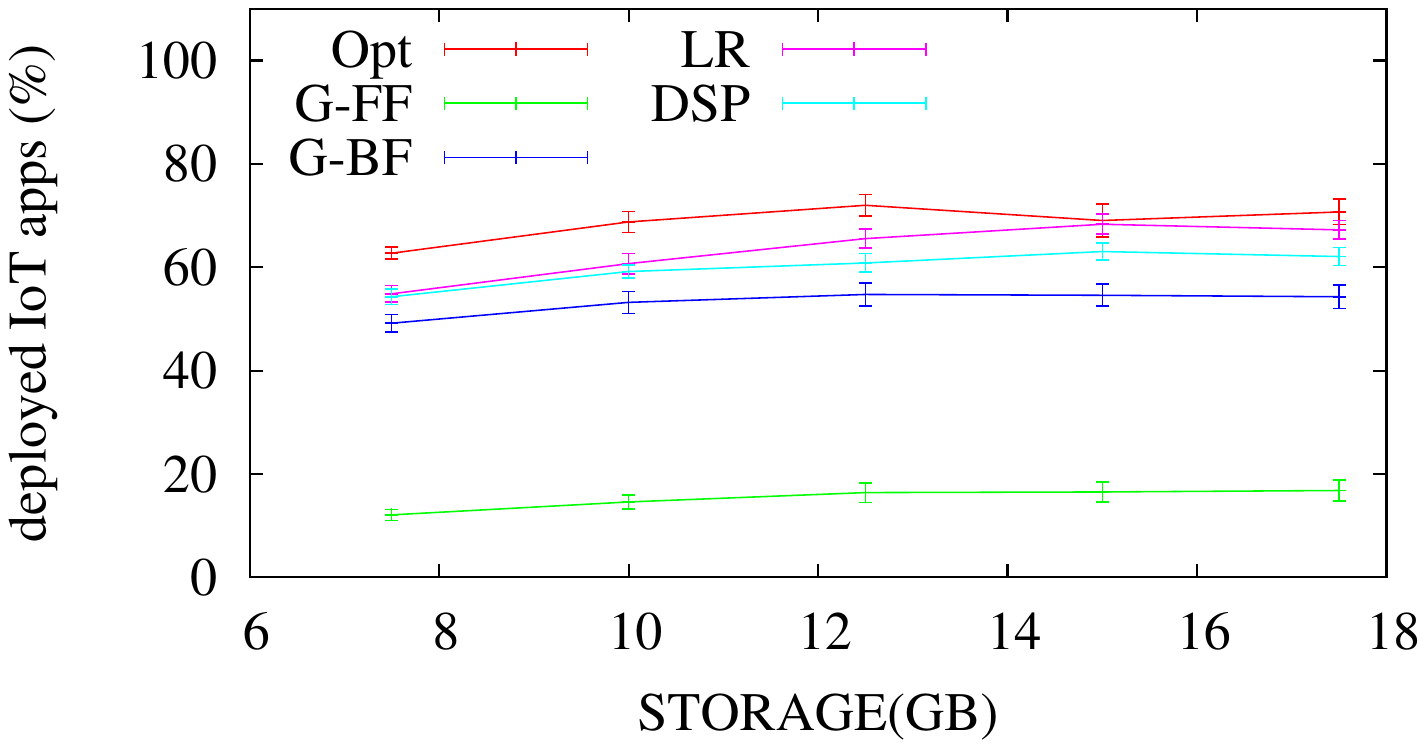}
    \label{fig:impact_sorage}
  }
 \subfloat[Fronthaul capacity]{
    \includegraphics[trim={2cm 4cm 1cm 4cm},clip,angle=0,width=0.45\columnwidth]{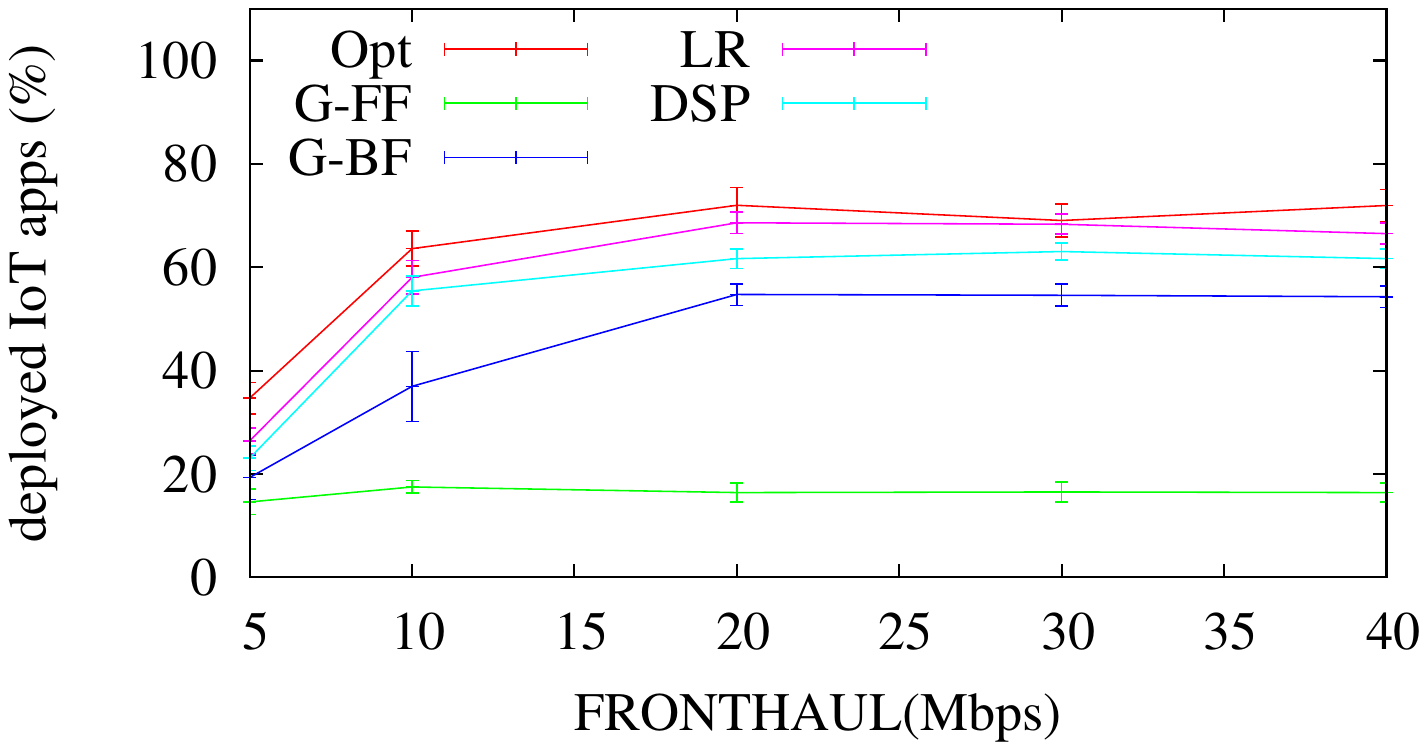}
    \label{fig:impact_fh}
  }
 \subfloat[Backhaul capacity]{
    \includegraphics[trim={2cm 4cm 1cm 4cm},clip,angle=0,width=0.45\columnwidth]{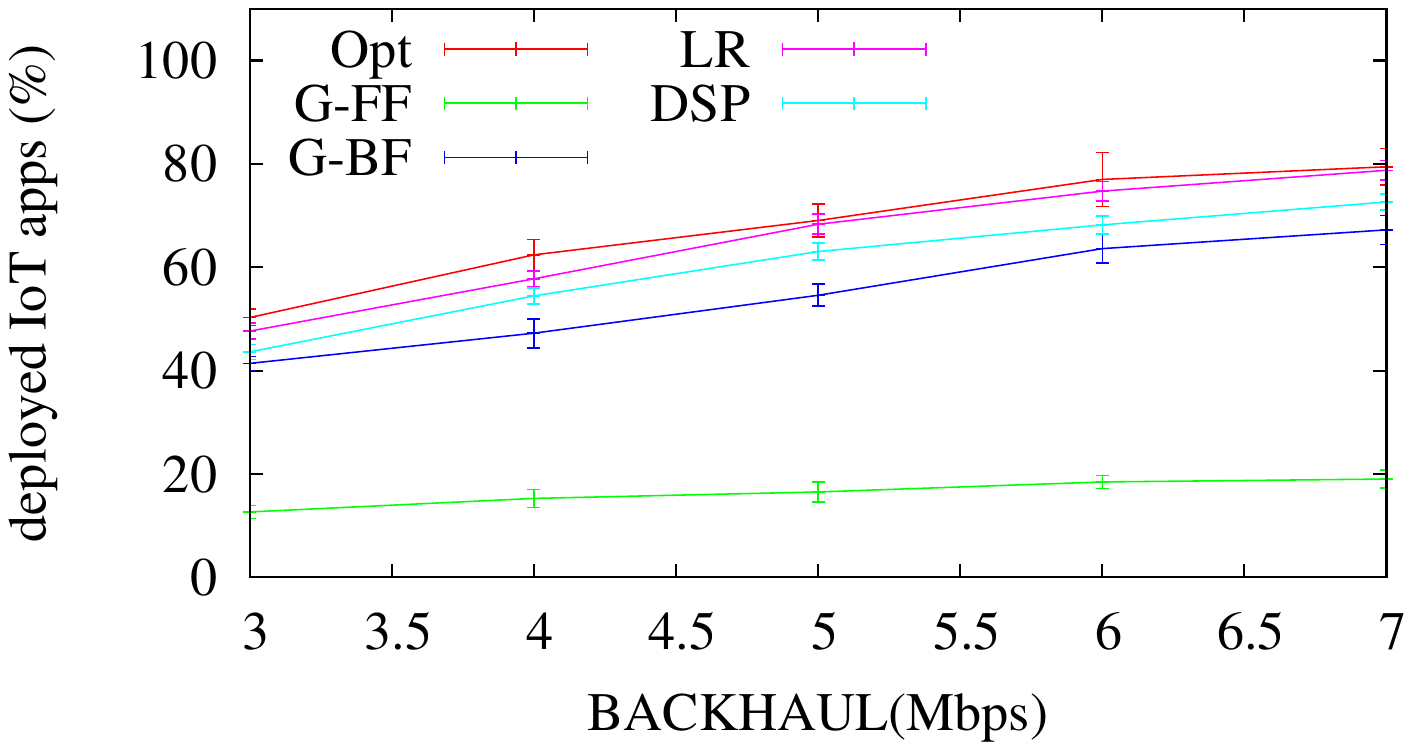}
    \label{fig:impact_bh}
  }
  \caption{Percentage of deployed applications for different~(a) computation, (b) memory, (c) fronthaul bandwidth, and (d) backhaul bandwidth capacities for $\alpha = 1.5$ and 60 service requests. \label{fig:impact}}
  \vspace{-0.25cm}
\end{figure*}

\noindent
In this section, we explore the impact of each resource on the efficiency of the allocation policies in terms of percentage of deployed services. Due to space limitations, we focus on the case $\alpha=0.6$ and 60 service requests.
In Figure~\ref{fig:impact_cpu}, We first explore in the impact of computation capacity $\Delta_m$ considering a range of values starting from 0.5GHz to 2.5GHZ. As expected, increasing computation capacity increases the number of deployed service slightly. In addition, the performance gap between the different policies remains stable in all combinations. Next, we show the impact of storage capacity $\Gamma_m$ in Figure~\ref{fig:impact_sorage}. The observed trends are similar to those depicted in Figure~\ref{fig:impact_cpu}, and again the effect of increasing storage capacity flattens out rapidly. More interesting results are shown in Figure~\ref{fig:impact_fh} and Figure~\ref{fig:impact_bh}, which depict the percentage of deployed services for different values of fronthaul ($B^{\uparrow}_h$) and backhaul ($B_{m_1m_2}$) bandwidth capacities, respectively. We can observe that an increase of fronthaul bandwidth capacity from 5Mps to 10 Mbps results in doubling the percentage of deployed services. This is due to the fact that many sensing resources should be activated around the PoIs, and a limited uplink bandwidth can prevent the activation of the needed sensors. However, a further increase of fronthaul bandwidth capacity has only a marginal impact on the efficiency of the proposed algorithms. On the contrary, the percentage of deployed service increases almost linearly with the backhaul bandwidth capacity. These results show that the amount of data exchanged between MEC hosts is non-negligible, and effective management of traffic flows within the MEC network eventually leads to superior performance. Finally, we conclude by noting that the gap between Opt and LR is small in all combinations\added{, while DSP is the second best performing scheme}, thus confirming the efficiency of the proposed heuristic.  

%

%
\subsection{Impact of service variability\label{sec:variability}}
\begin{figure}[th]
    \centering
    \includegraphics[trim={4cm 2cm 3cm 2cm},clip,angle=0,width=0.95\columnwidth]{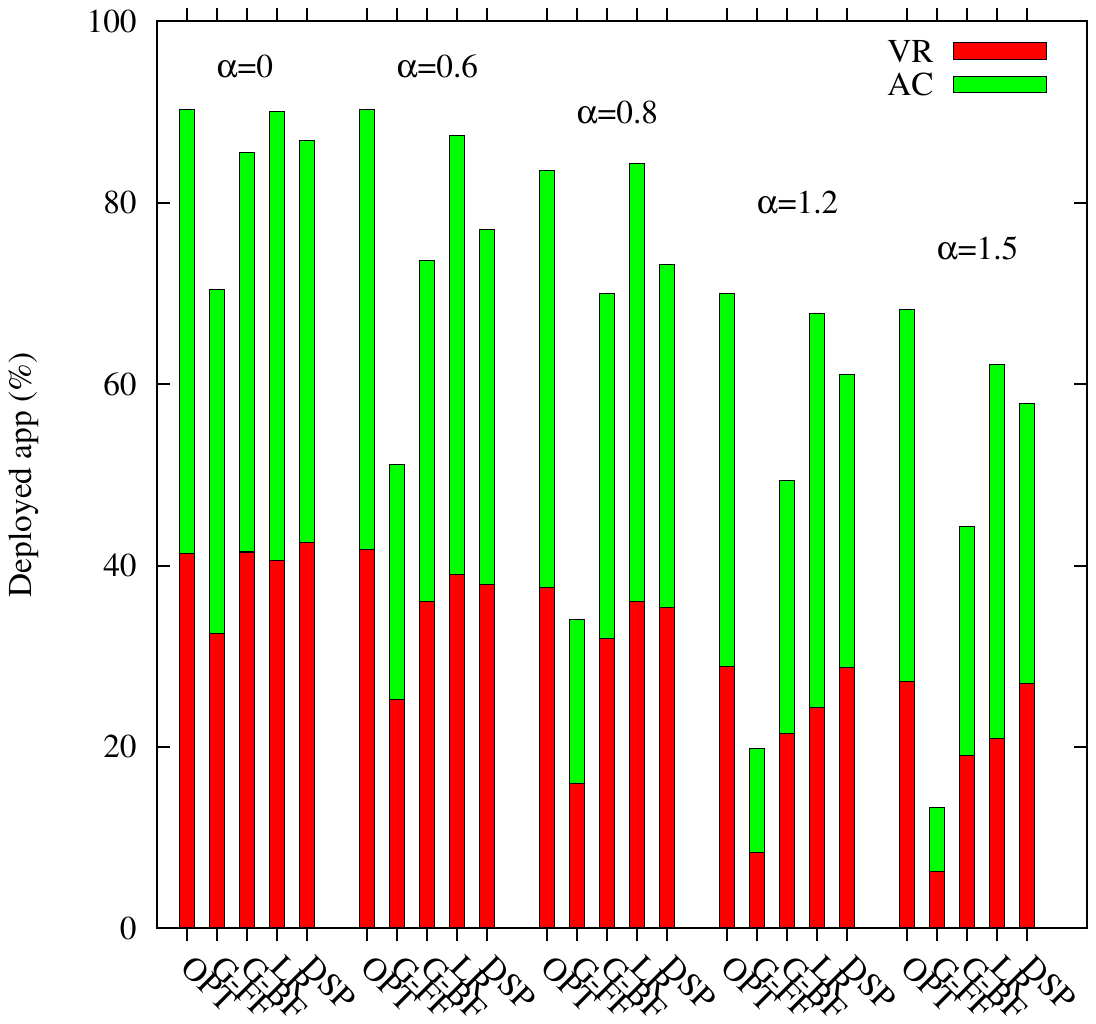}
    \caption{Distribution of fulfilled service requests across
the two types of services  for different values of zipf $\alpha$ parameter. The results are obtained with two PoIs and 50 service requests per class (\emph{Scen B}).}
    \label{fig:app_zipf_2class}
\end{figure}

\noindent
To conclude our evaluation, it is worth investigating which types of service requests are more efficiently handled by the proposed allocation policies. To this end, Figure~\ref{fig:app_zipf_2class} depicts the distribution of fulfilled service requests across the two types of services that we have previously defined, i.e. VR and AC, when there are 50 service requests per class and $\alpha$ value is varied from 0 to 1.5. Similarly to what observed in Figure~\ref{fig:app_zipf}, the total number of deployed applications decreases when $\alpha$ increases. However, AC services are less affected by the $\alpha$ value. For instance, when $\alpha=1.5$ 46\% of the VR services is discarded while only 18\% of AC services are discarded with Opt. In addition, Opt performs significantly better than the greedy-based solutions, with gains up  to  500\%  over  \added{G-FF}  and  up  to  54\%  over  \added{G-BF} (for $\alpha=1.5$). These performance gains are higher than those observed when only VR services are requested by end-users (see Figure~\ref{fig:app_zipf}). Finally the performance gap between Opt and LR is small, which confirms the efficiency of the heuristic also with heterogeneous service requests. 

To explain these results we can observe that AC services consume less resources than VR services. Thus, the objective function~(\ref{eq:opt_problem_MILP}) assigns a higher utility to AC services over VR services. Clearly, this depends on the adopted revenue model, which assign the same value to each type of service, independently of the resource usage (see equation~(\ref{eq:jr})). In addition, since AC services have a smaller resource footprint it is easier to distribute AC services in the MEC system to fully utilise the spare resources of individual MEC hosts. Opt and LR are more efficient than the greedy-based solutions in exploiting this multiplexing effect. \added{On the other hand, DSP only aims to minimise the network usage and it treats the two services equally.}  
%

%
\section{Related Work\label{sec:related}}
\noindent
There is a vast literature on resource management in edge computing environments, and MEC systems in particular, and the interested reader is referred to~\cite{2018_comst_edge_comm_survey,2019_csur_survey} for a comprehensive survey. Most of the earlier work on this research area focuses on the problem of optimal placement of computation tasks on suitable edge nodes based on the availability of edge resources and the demand patterns under static (e.g. in~\cite{2016_tnet_offloading}) or dynamic conditions (e.g. in~\cite{2018_jiot_offloading}). Recently, the problem of service placement and request routing in MEC systems where storage resource (i.e. caches) may be shared by requests of the same service from multiple end-users has received significant attention~\cite{2017_access_edge_survey}. For instance, the authors in~\cite{2018_mwc_cocaco} study the problem of computing task caching and formulate a MILP to minimise the task duration of all tasks under the caching capacity constraints. In~\cite{2018_tnet_offloading} the joint content caching, computing and bandwidth resources allocation problem is formulated as a MINLP problem to minimise the total energy consumption and network usage, and an ADMM-based distributed algorithm is proposed to solve it. The work in~\cite{2018_icds_edge_share} considered a similar problem as~\cite{2018_tnet_offloading} under the assumption that the coverage regions of the base stations are non-overlapping. A recent work extended these problem formulations to also account for services that need to pre-store a non-trivial amount of data, or consume significant downlink bandwidth to send the outcome of the processing task to the requesting user~\cite{2020_tnet_mec}. However, these works did not consider that multiple data streams generated by dispersed devices may need to be shared among many services with heterogeneous requirements through data caches. In addition, these studies overlooked the data dependencies when placing services, e.g. in terms of minimising the flow of data exchanged.   

Another thread of research related to this work are the studies that extend MEC capabilities to efficiently support IoT deployments and IoT applications~\cite{2018_comst_mec_iot}. One possible approach is to embed MEC-server functionalities within the conventional IoT gateways to allow local execution of IoT applications~\cite{2016_mce_iot_mec}. An alternative architecture is proposed in~\cite{2019_mcomstd_mec_iot}, where an additional MEC component, called MEC IoT platform, is introduced to virtualise IoT-gateway functionalities and to enable other MEC applications to interact with deployed IoT devices, hiding the complexity and heterogeneity of underlying IoT physical infrastructures. A network slicing framework for IoT services in hybrid architectures with MEC and centralised cloud systems is proposed in~\cite{2019_globecom_mec_iot}, which is based on a series of micro-services encapsulated as Virtual Network Functions (VNFs). SDN and VNF technligies are also used in~\cite{2018_jcna_iot_gateway} to \added{develop} a virtualised MEC platform for IoT applicatiuons. The authors in~\cite{2107_globecom_slaas} propose a new architecture comprising a central IoT Broker, which interacts with IoT applications and shapes IoT traffic to meet QoS requirements at application level, and a 5G Network Slice Broker in charge of negotiating with each IoT Broker network slice resources. A framework that allows uncoordinated access of IoT clients to serverless micro-services in MEC systems is developed in~\cite{2020_comnet_uncoordinated}. While our system is inspired by these works, our solution allows IoT tenants to have a better control of resource allocation during IoT service deployment.

\added{There are also many similarities between our scenario and the body of work that addresses the problem of resource allocation for data stream processing (DSP) applications, namely applications that can process unbounded streams of data in a near real-time fashion~\cite{2019_access_surveyDSP}. Typically, DSP applications can be decomposed into components that execute different processing tasks. Thus, they can be represented as a network of operators that connect data sources with data sinks (i.e., receivers of results). In this context, a classical problem is deciding on how to deploy these components on computing nodes in order to fulfil application requirements and considering computing and communication constraints. The reader is referred to ~\cite{2019_access_surveyDSP,2018_JCNA_DSP_edge_survey} for a comprehensive survey on DSP frameworks, while in this overview we are concerned with solutions that focus on edge/forg environments. For instance, the authors in~\cite{2020_deSouza_opt_model_DSP} formulated an \emph{assignment problem} as a MILP to map DSP operators onto available resources in a mixed cloud-edge infrastructure, while jointly minimising task response time and deployment costs. The proposed model also allows to parallelise the execution of operators onto multiple computing resources. To address the scalability issue of the model in~\cite{2020_deSouza_opt_model_DSP}, in a follow-up papers the authors propose a resource selection technique to reduce the state space of the problem~\cite{2020_deSouza_fast_pruning_DSP}. Two greedy strategies are described in~\cite{2019_smartcomp_DSP_greedy} to solve the operator assignment problem, which prioritise the allocation of resources close to the data sources. A MILP formulation of the operator assignment problem is also proposed in~\cite{2019_TPDS_ODP_best} aiming to jointly minimise the expected application response and unavailability, as well as network usage. Then, a set of model-free and model-based heuristics are proposed to find a solution. Recently, the authors of~\cite{2020_DSPMEC} have formulated the operator placement problem for DSP applications in IoT environments. Our work differs from these solutions because we consider consider the possibility of sharing data flows among processing applications using edge caches. Furthermore, most of the problem formulations assumes that the mapping between data sources and operators is fixed, while we allow a dynamic assignment to optimise the routing of data flows within the network of edge devices.} 
\vspace{-0.5cm}

%
\section{Conclusions\label{sec:conclusions}}
\noindent
In this paper, we have studied the problem of joint service placement and data management in mobile edge computing systems. We have targeted a scenario where an IoT service provider leverages network, computing and storage resources of a MEC-enabled system to fulfil the heterogeneous requirements of delay-sensitive and data-intensive IoT applications. The proposed approach aims at the maximisation of the revenue due to the application deployment process (in terms of number of deployed applications) while minimising the cost of the required physical resources. Furthermore, our model accounts for 
data dependencies when placing services, e.g. in terms of minimising the 
flow of data exchanged between edge servers. We have developed both an optimisation framework and a heuristic algorithm. Finally, we have proposed a MEC architectural solution to enable the IoT service provider to transparently interact with the MEC sytem. The results obtained with realistic scenarios show the efficiency of the proposed approach, especially when traffic demands generated by the service requests are not uniform. Interesting directions for future work include exploring the cooperation and competition between multiple IoTSPs using the same MEC systems, as well as investigating general MEC topology and different data models. 

%
\section*{Acknowledgement\label{sec:ack}}
\noindent
This work is partly funded by the EC under the H2020 INFRADEV-2019-3 SLICES-DS (951850) and the H2020-EU.2.1.1 MARVEL (957337) projects, and by the Italian Ministry of Economic Development under the ARTES 4.0 project. 

\bibliography{ref.bib}

\end{document}